\documentclass[conference]{IEEEtran}
\IEEEoverridecommandlockouts
\usepackage{cite}
\usepackage{amsmath,empheq,amssymb,amsfonts}
\usepackage{textcomp}
\usepackage{graphicx}
\usepackage{dirtytalk}
\usepackage{mathrsfs}
\usepackage[table]{xcolor}
\usepackage{tikz}
\usepackage{neuralnetwork}
\usepackage{subfig}
\usepackage{booktabs}
\usepackage[ruled,vlined,linesnumbered]{algorithm2e}
\usepackage{setspace}
\usepackage{url}
\usepackage{enumerate}
\usepackage{diagbox}
\usepackage{rotating}
\usepackage{float}
\usepackage[bookmarks=false]{hyperref}
\usepackage{comment}
\usepackage{multirow}
\usepackage{nopageno}
\usepackage{enumitem}
\usepackage[normalem]{ulem}
\usepackage{soul}
\usepackage{balance}
\usepackage{pifont}
\usepackage{ifthen}
\usepackage{balance}
\usepackage{pifont}
\newcommand{\cmark}{\textcolor{green}{\ding{51}}}   
\newcommand{\xmark}{\textcolor{red}{\ding{55}}}   

\usepackage[colorinlistoftodos,prependcaption, textsize=small, textwidth=30]{todonotes}

\usepackage{algpseudocode}
\usepackage{threeparttable} 

\usepackage{amstext}

\def\BibTeX{{\rm B\kern-.05em{\sc i\kern-.025em b}\kern-.08em
    T\kern-.1667em\lower.7ex\hbox{E}\kern-.125emX}}
\begin{document}

\title{
Benchmarking RL-Enhanced Spatial Indices Against Traditional, Advanced, and Learned Counterparts}

\author{
\IEEEauthorblockN{Guanli Liu}
\IEEEauthorblockA{
\textit{The University of Melbourne}\\
guanli.liu1@unimelb.edu.au}
\and
\IEEEauthorblockN{Renata Borovica-Gajic}
\IEEEauthorblockA{
\textit{The University of Melbourne}\\
renata.borovica@unimelb.edu.au}
\and
\IEEEauthorblockN{Hai Lan}
\IEEEauthorblockA{
\textit{The University of Queensland}\\
h.lan@uq.edu.au}
\and
\IEEEauthorblockN{Zhifeng Bao}
\IEEEauthorblockA{
\textit{The University of Queensland}\\
zhifeng.bao@uq.edu.au}
}

\newtheorem{definition}{Definition}
\newtheorem{lemma}{Lemma}
\newtheorem{theorem}{Theorem}
\newtheorem{corollary}{Corollary}

\newtheorem{example}{Example}

\newcommand\rtree{R-tree}
\newcommand\rstar{R$^*$-tree}
\newcommand\rlrtree{RLR-tree}
\newcommand\platon{PLATON}
\newcommand\kdtree{KD-tree}
\newcommand\kdbtree{KDB-tree}
\newcommand\kdgreedy{GKD-tree}
\newcommand\qdtree{Qd-tree}
\newcommand\waffle{Waffle}
\newcommand\zrtree{ZR-tree}
\newcommand\zrrtree{ZRR-tree}
\newcommand\bmtree{BM-tree}
\newcommand\zmindex{ZM-index}
\newcommand\lisa{LISA}

\newcounter{observation}  

\setcounter{observation}{0}  %







\setlength\abovecaptionskip{0pt}
\setlength\belowcaptionskip{-2pt}
\setlength{\floatsep}{1pt plus 0.5pt minus 1pt}
\setlength{\textfloatsep}{1pt plus 0.5pt minus 1pt}
\setlength{\intextsep}{1pt plus 0.5pt minus 1pt}

\maketitle

\begin{abstract}
Reinforcement learning has recently been used to enhance index structures, giving rise to reinforcement learning-enhanced spatial indices (RLESIs) that aim to improve query efficiency during index construction. However, their practical benefits remain unclear due to the lack of unified implementations and comprehensive evaluations, especially in disk-based settings.

We present the first modular and extensible benchmark for RLESIs. Built on top of an existing spatial index library, our framework decouples index training from building, supports parameter tuning, and enables consistent comparison with traditional, advanced, and learned spatial indices.

We evaluate 12 representative spatial indices across six datasets and diverse workloads, including point, range, $k$NN, spatial join, and mixed read/write queries. Using latency, I/O, and index statistics as metrics, we find that while RLESIs can reduce query latency with tuning, they consistently underperform learned spatial indices and advanced variants in both query efficiency and index build cost.
These findings highlight that although RLESIs offer promising architectural compatibility, their high tuning costs and limited generalization hinder practical adoption.
\end{abstract}


\section{Introduction}\label{sec:introduction}
The explosive growth of location-based services and spatial analytics has led to a surge in spatial data, often far exceeding the capacity of main memory. To ensure scalable query processing, traditional disk-based spatial indices have long been used in production systems~\cite{PostGIS,MS_SQL,Oracle}. However, increasing dataset sizes, complex workloads, and data skew present growing challenges to these traditional structures, particularly in maintaining high query performance across diverse scenarios.

Recent work has explored \textit{learning-based spatial indexing methods}, which leverage machine learning or reinforcement learning (RL) to enhance index construction and query efficiency~\cite{ZM,RSMI,Flood,LISA,BMTree,qdtree,RLR-tree,PLATON}. These methods can be broadly categorized based on their integration of learning models:
(1) \textbf{\underline{L}earned \underline{S}patial \underline{I}ndices (LSIs)}, such as \zmindex~\cite{ZM}, RSMI~\cite{RSMI}, Flood~\cite{Flood}, and \lisa~\cite{LISA}, employ machine learning models to construct entirely new index structures that directly capture the spatial data distribution. These indices require custom query and update mechanisms.
(2) \textbf{\underline{RL}-\underline{E}nhanced \underline{S}patial \underline{I}ndices (RLESIs)}, such as \bmtree~\cite{BMTree}, \qdtree~\cite{qdtree}, and \rlrtree~\cite{RLR-tree}, retain traditional structures and use RL models to optimize decision-making during index building, such as determining partition boundaries.

While both LSIs and RLESIs aim to accelerate spatial queries, they differ fundamentally in design philosophy: LSIs propose entirely new structures, whereas RLESIs optimize within existing ones. This distinction gives RLESIs a practical advantage in terms of deployability and ease of integration into existing systems.

Despite their promise, RLESIs remain insufficiently studied along two key dimensions: their \textit{external} performance relative to diverse and competitive baselines, and their \textit{internal} sensitivity to parameter choices. Existing evaluations are often conducted in isolated settings with fixed parameters and narrow baselines~\cite{BMTree,qdtree,RLR-tree}, limiting generalizable insights.

This raises two key research questions:
\textit{(1) How do RLESIs compare to traditional, advanced, and learned spatial indices across diverse workloads?}
\textit{(2) How do configuration parameters, such as training sample size and tuning frequency, affect the query performance of RLESIs?}
Answering these questions is non-trivial due to the lack of a unified evaluation framework. 

To address \textit{Question 1}, we develop the first unified benchmarking framework for evaluating RLESIs alongside traditional, advanced, and learned spatial indices. Our framework supports end-to-end evaluation by decoupling index learning from index building, enabling reusable implementations and consistent comparisons across indexing methods.

Existing benchmarks~\cite{sosd,ULI_Evaluation,are_learned_ready,Benchmark_ls} focus on one-dimensional indices~\cite{ALEX,btree,fiting_tree,pgm,RMI,LIPP,RadixSpline}, which differ significantly from RLESIs in terms of dimensionality and query types. While a benchmark for learned spatial indices exists~\cite{learnedbench}, it targets in-memory LSIs and is not suitable for disk-based RLESIs. Additionally, fragmented implementations, e.g., \rlrtree~\cite{RLR-tree} using standalone R-trees and \bmtree~\cite{BMTree} relying on external environments, complicate fair evaluation.

Our benchmarking framework includes representative indices from three major spatial indexing paradigms: data partitioning, space partitioning, and mapping. These span traditional, learned, and RL-enhanced variants. We prioritize open-source implementations to ensure reproducibility.

\begin{figure}[h]
  \centering
 \includegraphics[width=0.52\textwidth]{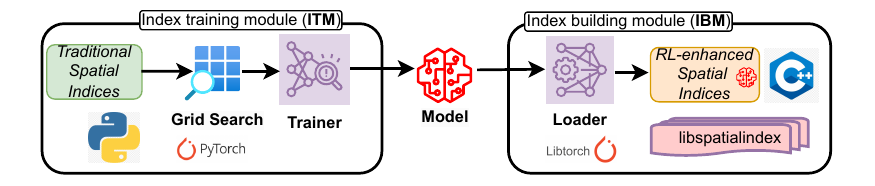}
  \caption{An overview of the benchmarking framework.}
  \label{fig:overview_framework}
\end{figure}

As illustrated in Figure~\ref{fig:overview_framework},
the framework comprises two primary modules: the \textbf{Index Training Module (ITM)} and the \textbf{Index Building Module (IBM)}:
\begin{enumerate}[leftmargin=4mm,label=\textbullet]
    \item \textbf{ITM} provides a standardized training environment for RLESIs through a \textit{trainer} built on PyTorch~\cite{pytorch}, ensuring consistency in training, parameter tuning, and RL model generation. PyTorch is chosen for its compatibility with existing RLESI baselines, which are predominantly implemented in Python.

    \item \textbf{IBM} extends the disk-based indexing library \textit{libspatialindex}~\cite{libspatialindex} to support RLESIs, enabling the seamless integration into spatial systems. A core component of IBM is the \textit{loader}, which leverages PyTorch’s C++ API to load trained RL models produced by ITM, enabling integration with traditional disk-based indices.
\end{enumerate}

To address \textit{Question 2}, we incorporate a structured \textit{grid search} mechanism within ITM for parameter tuning. This component systematically explores configuration spaces to minimize query costs while balancing index building overhead and query performance. Grid search is tailored for each dataset and workload, and constrained by a training cost threshold to ensure practical feasibility.

This paper makes the following key contributions:

\noindent\textbf{\ding{192} A unified benchmark for comprehensive evaluation of RLESIs.}  
We present the first systematic evaluation of RL-enhanced spatial indices, comparing them with traditional, advanced, and learned baselines under diverse workloads.  Our benchmark isolates learning effects from structural differences and fine-tuning, and quantifies RLESIs under diverse workloads. All components are released as open source\footnote{\url{https://github.com/Liuguanli/rl_spatial_benchmark/tree/icde26}} to facilitate reproducibility and future extensions.

\noindent\textbf{\ding{193} A comprehensive and reusable evaluation setup for benchmarking spatial indices.}  
Our framework includes diverse datasets, workloads (point, range, and $k$NN queries), and metrics (latency at P50 and P99, I/O operations, index-level statistics) to reflect realistic spatial scenarios.

\noindent\textbf{\noindent\ding{194} 
Our extensive experiments yield several key observations (O):}
\textbf{(1) Query Performance (O1-O5):}
RLESIs provide modest gains over traditional indices but often underperform compared to advanced and learned baselines.
\textbf{(2) Query Variation (O6-O8):}
RLESIs exhibit high latency for range and $k$NN queries under varying query parameters, except for \bmtree~\cite{BMTree}.
\textbf{(3) Tail Latency (O9-O11):} 
RLESIs generally show increasing latency from P1 to P99, \rlrtree\ and \qdtree\ suffer sharp spikes at high percentiles (e.g., P90 to P99).
\textbf{(4) Insertion Performance (O12-O14):}
Insertion latency is high for RLESIs, with \bmtree\ being the worst; however, they maintain query performance under heavy insertions.   
\textbf{(5) Index Build (O15-O18):}
RLESIs incur high index build costs due to model training overhead. 
Additionally, data partitioning-based indices incur higher build costs due to frequent I/O operations, while mapping-based indices are more space-efficient.
\textbf{(6) Cardinality Variation (O19):} 
Most indices, including RLESIs, maintain performance scalability across dataset sizes and query types.
\textbf{(7) Parameter Tuning (O20-O22):}
Proper parameter optimization of RLESIs can lead to up to 120$\times$ reductions in query latency.

\noindent\textbf{\ding{195} We systematically evaluate RLESIs against other baselines and identify challenges and limitations of RLESIs, e.g., high training costs.}
Our study reveals that while RLESIs can outperform traditional indices in some cases, they suffer from high training costs and require careful parameter tuning. 
To mitigate training costs, we adopt advanced reward functions~\cite{LBMC}, achieving up to 27\% training time reduction for \bmtree\ without compromising query performance.

\section{Spatial Queries and Indices Revisited}\label{sec:spatial_indices}
This section examines spatial queries and categorizes spatial indices into three primary types: data partitioning, space partitioning, and mapping~\cite{ learned_spatial_indexes, Flood, learnedmultiindexNeurIPS}, based on their build methods. In each category, we analyze its
traditional, advanced, learned (LSI), and RL-enhanced (RLESI) versions.
In this paper, we focus on points as the primary spatial data type because RLESIs are optimized for handling points.

\vspace{-0.5em}
\subsection{Spatial Queries}
Spatial queries enable the analysis and retrieval of spatial information based on the relationships and properties of geographic entities. 
Common query types include point queries, range queries, $k$ nearest neighbor ($k$NN) queries, and spatial joins, each critical for spatial analysis.

\textit{Point queries} retrieve the points that exactly match a specified query point. They can be viewed as a special case of range queries, where the query range is limited to a single point. A point query is formally defined as follows:

\begin{definition}[Point Query (PQ)]
    Given a point $q \in E^d$, find all points $p$ in dataset $D$ that are equal to $q$: 
    \begin{equation}
\small
        PQ(q) = \{p \in D \mid p = q\}.
    \end{equation}
\end{definition}

\textit{Range queries} retrieve all points within a specified query range, also known as a query window, which is represented as a (hyper)rectangle. A range query is defined as follows:

\begin{definition}[Range Query (RQ)]
    Given a $d$-dimensional interval $I^d = [l_1,u_1] \times [l_2,u_2] \times \dots \times [l_d,u_d]$, find all points $p$ in dataset $D$ that are contained within $I^d$:
    \begin{equation}
\small
        RQ(I^d) = \{p \in D \mid p \in I^d\}.
    \end{equation}
\end{definition}

\textit{$k$NN queries} generalize the nearest neighbor (NN) query (i.e., $k=1$) by returning the $k$ points in the dataset that are ``nearest'' to a given query point, as defined by a distance function $dist$. A $k$NN query is formally defined as follows:

\begin{definition}[$k$ Nearest-Neighbor Query ($k$NNQ)]
    Given a query point $q$ and a distance function $dist$, return all points $p$ in subset $S$ ($|S|=k$) that have the minimum distance to $q$:
    \begin{equation}
\small
        kNNQ(k, q) = \{p \in S \mid \forall p' \in D \setminus S: \text{dist}(p, q) \leq \text{dist}(p', q)\}.
    \end{equation}
\end{definition}

\textit{Spatial joins} identify all pairs of points from two spatial datasets that satisfy a specified spatial predicate $\theta$~\cite{R_tree_join,Rtree_book}. Given two spatial datasets $R$ and $S$, a spatial join query returns all pairs ($o$, $o^{'}$), where $o \in R$ and $o{'} \in S$ that satisfy a spatial predicate $\theta$. Common predicates $\theta$ include \texttt{intersects}, \texttt{contains}, and \texttt{distance}.

\begin{definition}[Spatial Join (SJ)]
Given two spatial datasets $R$ and $S$, and a spatial predicate $\theta$, return all pairs of objects $(o, o')$ such that:
\begin{equation}
\small
SJ(R, S, \theta) = \{ (o, o') \mid o \in R, o' \in S, \theta(o, o') = \text{true} \}.
\end{equation}
\end{definition}

\vspace{-0.5em}
\subsection{Classification of Spatial Indices}\label{subsec:classification_spatial_indices}
Following the classification outlined in \cite{CaseSpatial,RSMI}, we categorize spatial indices into three primary categories: data partitioning-based, space partitioning-based, and mapping-based spatial indices.

\subsubsection{\underline{D}ata \underline{P}artitioning-based (DP-based)}
DP-based spatial indices partition data into smaller regions with minimal overlap to reduce query search space. By limiting overlap, these indices enable efficient query processing by retrieving fewer tree branches and nodes, leading to faster response times.

\textbf{Traditional:}
\emph{\rtree s}~\cite{rtree} are widely used for spatial indexing~\cite{PostGIS}.
An \rtree\
organizes data into a tree-structured nested rectangle, where each tree node represents a minimum bounding rectangle (MBR) that contains child nodes or data points.
When a node exceeds its capacity, it is split into two smaller nodes based on criteria such as spatial area or margin, aiming to reduce query costs.

\textbf{Advanced:}
Advanced versions of R-tree~\cite{rtree}, such as \rstar~\cite{r_star_tree}, revised \rstar~\cite{revised_r_star_tree}, and PR-tree~\cite{PR_tree}, 
focus on reducing node overlap and optimizing node splitting to improve query performance.
\rstar\ minimizes the total area of MBRs and their overlap through a complex node-splitting strategy and reinsertion mechanism.
STR-tree~\cite{STR} avoids overlap through recursive sorting and tiling, efficiently partitioning data into spatially coherent groups. 

\textbf{Learned:}
AI+R-tree~\cite{AI_RTREE} integrates ML models to enhance traditional R-trees by directly predicting the target leaf node for queries, which avoids retrieving multiple leaf nodes.
\platon~\cite{PLATON}, despite incorporating RL, is categorized as an LSI because it introduces a novel index structure.
It uses a learned partitioning policy based on packing R-tree tailored to specific data and query workloads.

\begin{figure}[h]
  \centering
  \includegraphics[width=0.45\textwidth]{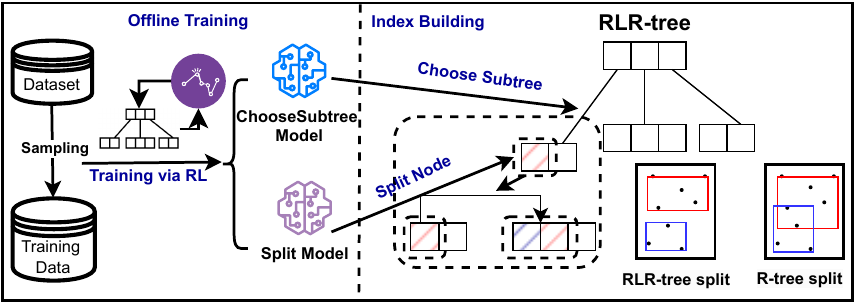}
  \caption{An overview of \rlrtree\ build.}
  \label{fig:rlrtree_demo}
\end{figure}

\textbf{RL-enhanced:}
ACR-tree~\cite{ACR-Tree} and
\rlrtree~\cite{RLR-tree} refine the \rtree~\cite{rtree} structure rather than introducing new ones.
As shown in Figure~\ref{fig:rlrtree_demo},
\rlrtree~\cite{RLR-tree} includes an offline training phase using sampled datasets.
The trained models are then integrated into the index build phase of the \rtree~\cite{rtree}.
During data insertion, \rlrtree~\cite{RLR-tree} leverages the models to select subtrees and predict node splits when the target node is full.
ACR-tree~\cite{ACR-Tree} employs 
an Actor-Critic~\cite{Actor_critic} RL method, where the Critic estimates the long-term cost, and the Actor determines actions like splitting or packing.

\subsubsection{\underline{S}pace \underline{P}artitioning-based (SP-based)}
SP-based indices partition the entire space into smaller, non-overlapping regions, typically through a recursive approach.
Unlike data partitioning, where overlaps can occur between the MBRs of partitioned data objects, space partitioning ensures no overlaps exist.
Space partitioning continues until the spatial objects within an area are sufficiently reduced to fit into a leaf node. The design of these indices aims to reduce the number of comparisons needed to locate a spatial object.

\textbf{Traditional:}
\emph{\kdtree}~\cite{kdtree}, \emph{Quadtrees}~\cite{quadtree}, and \textit{Grid-Files}~\cite{grid} are typical examples. \kdtree~\cite{kdtree} partitions a multi-dimensional space by alternately choosing one dimension and finding the median to divide the space into two areas. 
Quadtree~\cite{quadtree} recursively partitions a two-dimensional space into four quadrants until the contained data in each quadrant meets certain criteria.
In contrast, \textit{Grid-File}~\cite{grid} partitions multidimensional space into a fixed grid of cells, wherein each cell links to a bucket that stores the records.

\textbf{Advanced:}
\kdbtree~\cite{KDB} merges the dimension-wise splitting of a \kdtree~\cite{kdtree} with the multi-way block storage of a B-tree~\cite{btree}.
Its inner nodes hold multiple partitions, optimizing data retrieval for disk-based storage. This approach reduces disk access, making it effective for large datasets.
\kdgreedy~\cite{qdtree} heuristically adjusts partitions based on query patterns to enhance query performance.

\begin{figure}[h]
  \centering
  \includegraphics[width=0.45\textwidth]{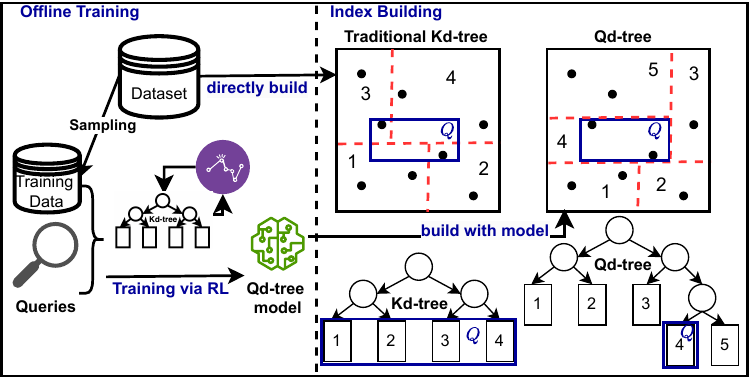}
  \caption{An overview of Qd-tree build.}
  \label{fig:qdtree_demo}
\end{figure}

\textbf{Learned:}
LISA~\cite{LISA} partitions the data space using a grid and maps data points based on a weighted aggregation of their coordinates. It utilizes learned functions to assign a shard ID to each point.
SPRIG~\cite{SPRIG} is similar to LISA~\cite{LISA} in using a grid for partitioning, but uses bi-linear interpolation functions for prediction.
Flood~\cite{Flood} and Tsunami~\cite{ tsunami} are memory-based, which partition a $d$-dimensional space using a 
($d$-1)-dimensional grid, indexing the points 
by their coordinates 
in the last dimension using RMI~\cite{RMI}.
CaseSpatial~\cite{CaseSpatial} and Spatial-LS~\cite{Spatial-LS} are in-memory indices leveraging different partition techniques.
Waffle~\cite{Waffle} is
a workload-aware spatial index 
that combines the space and data partitioning.

\textbf{RL-enhanced:}
The \textit{\qdtree}~\cite{qdtree} 
uses RL to guide the space partitioning process of a \kdtree~\cite{kdtree}, which helps minimize the number of data block accesses given a query workload (assuming that data points are grouped and stored in blocks). 
As shown in Figure~\ref{fig:qdtree_demo}, the 
\qdtree~\cite{qdtree} undergoes offline training 
with sampled datasets and queries to learn a partitioning strategy based on \kdgreedy~\cite{qdtree}. 
During each training episode, an intermediate \qdtree~\cite{qdtree} is incrementally built, making partitioning decisions until the tree is complete.

\subsubsection{\underline{M}a\underline{p}ping-based (MP-based)} 
MP-based indices map multi-dimensional data objects into one-dimensional values.
The key idea is to order spatial objects by their mapped values to ensure objects close in the multi-dimensional space remain close in the one-dimensional mapped values.
Once ordered, the data can be efficiently packed into data blocks and indexed using \textit{bulk-loading}, optimizing both storage and query performance.

\textbf{Traditional:}
Space-filling curves (SFCs) are widely used for mapping 
purposes due to their effectiveness in 
preserving the locality of spatial objects. 
The most used SFCs are
\emph{Z-curve}~\cite{Zcurve} and \emph{Hilbert-curve}~\cite{hilbert}, which map spatial coordinates into one-dimensional curve values.
These mappings support the build of spatial indices, namely \textit{\zrtree}~\cite{ZRtree} and \textit{HR-tree}~\cite{HRtree}. 
The use of SFCs ensures that objects within a node are likely to be in proximity, thereby minimizing overlap and enhancing query performance.

\textbf{Advanced:} 
A key challenge in traditional SFCs is selecting an appropriate bit resolution.
Low bit resolutions lead to excessive data clustering, while high resolutions result in overly fine-grained partitions, adding unnecessary complexity.
To address this, \textit{rank-space}~\cite{rank_space} is proposed with a worst-case query guarantee.
Instead of using actual values, this method 
calculates the curve value based on the rank of each dimension value. This approach can minimize the use of bits while ensuring each spatial object has a unique curve value. 
\zrtree~\cite{ZRtree} that integrates rank-space is called \zrrtree~\cite{rank_space}.

\textbf{Learned:}
Learned MP-based indices leverage machine learning for improved mapping and indexing.
\textit{ZM-index}~\cite{ZM} maps data points according to their Z-curve~\cite{Zcurve} values for sorting, followed by indexing using Recursive Model Indexing (RMI)~\cite{RMI}.
ML-Index\cite{ML-Index} utilizes RMI for indexing and incorporates the \emph{iDistance} technique\cite{idistancetods} to map data points to one-dimensional values.
RSMI~\cite{RSMI} employs SFCs to create a hierarchy of space partitions, mapping data points to their corresponding partition IDs. 

\begin{figure}[h]
\centering
\subfloat[\bmtree~\label{fig:bmtree_1}]   {
    \includegraphics[width=0.25\textwidth]{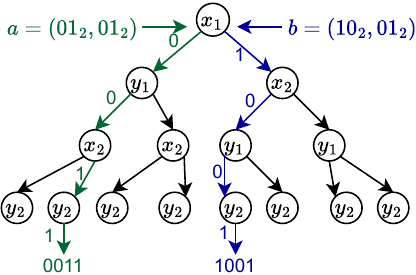}
}
\subfloat[A piecewise SFC~\label{fig:bmtree_2}]   {
\includegraphics[width=0.15\textwidth]{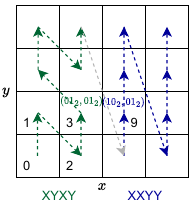}
}
\caption{An example of a piecewise SFC (\bmtree).}\label{fig:bmtree}
\end{figure}

\begin{figure*}
  \centering
  \includegraphics[width=1.00\textwidth]{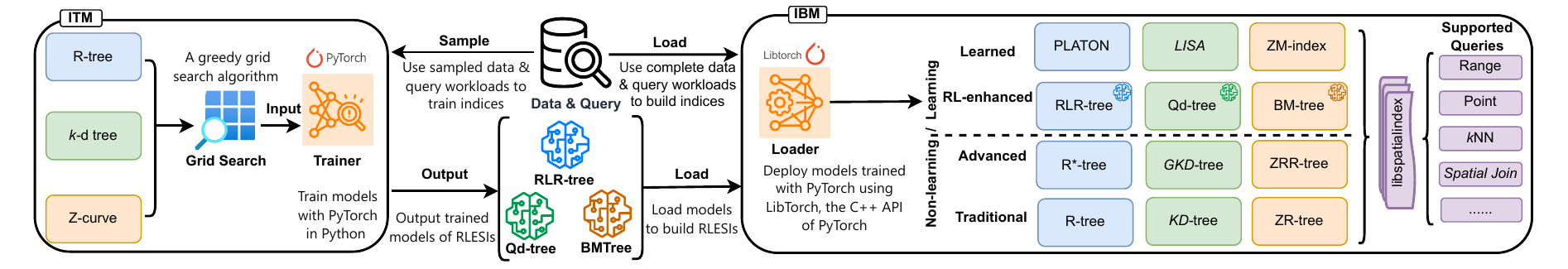}
  \caption{Framework for benchmarking spatial indices: learning, deployment, and build in libspatialindex.}

  \label{fig:rl_spatial_benchmark}
   \vspace{-1.5em}
\end{figure*}

\textbf{RL-enhanced:}
\bmtree~\cite{BMTree} and LMSFC~\cite{LMSFC} both learn space-filling curves (SFCs). LMSFC generates a global SFC via Bayesian optimization~\cite{SMBO}, while \bmtree\ constructs piecewise SFCs by partitioning the data space with a quadtree~\cite{quadtree}, assigning each partition a specific SFC segment. 
Like \qdtree~\cite{qdtree} and \rlrtree~\cite{RLR-tree}, \bmtree\ requires offline training, whereas its output is not an RL model but a tree structure that encodes the learned SFC.
In Figure~\ref{fig:bmtree_1}, the bits for $a = (01_2, 01_2)$ are $x_1=0$, $x_2=1$, $y_1=0$, and $y_2=1$. Traversing \bmtree\ from the root, one follows left for 0 and right for 1, resulting in a curve value of 3 for $a$ (Figure~\ref{fig:bmtree_2}).

\section{Benchmarking Framework}\label{sec:framework}

Our benchmarking framework comprises two modules: 
index training module (ITM) and index building module (IBM). 
This design is more effective than a standalone module, as it avoids the complexity of integrating training (often in Python) and implementation (typically in C++) in a single pipeline, and also enables the reuse of trained RLESIs.

In Figure~\ref{fig:rl_spatial_benchmark}, 
ITM is a Python-based module designed to train RLESIs by selecting best-performing parameters.
For each RLESI, we use configurations specific to the dataset and workload, testing various parameter combinations to find the best-performing setup. 
The traditional version, such as \rtree~\cite{rtree}, 
serves as the input to the \textit{trainer}.
After an RLESI is trained, the trainer can output the trained models. 
For \rlrtree~\cite{RLR-tree} and \qdtree~\cite{qdtree}, their outputs are the PyTorch~\cite{pytorch} models, while for \bmtree~\cite{BMTree}, the output is a tree structure that represents a piecewise SFC. 

IBM supports disk-based spatial indices and implements them in C++.
It extends from a well-known spatial index library libspatialindex~\cite{libspatialindex}, which already supports {\rtree s}~\cite{rtree}, {\rstar s}~\cite{r_star_tree}, and the skeleton of MP-based indices.
Building on this foundation, our framework supports point, range, and $k$NN queries, as well as spatial joins. 
To integrate trained models from ITM, we incorporate a  \textit{loader} component using \textit{LibTorch}, the C++ API of PyTorch~\cite{pytorch}.
LibTorch is only used for static model loading.
The loader takes PyTorch models and adapts them for C++ programs, which ensures seamless integration into the indexing pipeline.

\begin{algorithm}[h]
\setstretch{1.0}
    \begin{small}
      \caption{Tuning of RLESI} \label{alg:efficient_search}
      \KwIn{Parameter space $\mathcal{P}$, build cost limit $T_{build}$}
      \KwOut{Best-performing  parameter configuration $\mathcal{P}_{opt}$}
      $min\_query\_cost \leftarrow \infty$, $\mathcal{P}_{opt} \leftarrow \text{null}$\;
      
    /* $\mathcal{P}$:
      \{ ``epoch": [8, ..., 12], ``sample\_size": [1k, ..., 16k] \}*/

      \For{$\mathbf{p} \in \texttt{generate\_configurations}(\mathcal{P})$} {
          \If{\texttt{is\_build\_required($\mathbf{p}$, $\mathcal{P}_{opt}$)}} {
              $build\_cost, index \leftarrow \texttt{train\_index}(\mathbf{p})$\;
              \If{$build\_cost < T_{build}$} {
                  $query\_cost \leftarrow \texttt{run\_queries}(index)$\;
                  \If{$query\_cost < min\_query\_cost$} {
                      $min\_query\_cost \leftarrow query\_cost$, $\mathcal{P}_{opt} \leftarrow \mathbf{p}$\;
                  }
              }
            }
      }
      \Return $\mathcal{P}_{opt}$\;
    \end{small}
\end{algorithm}

\subsection{Implementation of Parameter Tuning}\label{subsec:tuning}
Our benchmark employs a structured grid search process for (hyper)parameter tuning of RLESIs, as outlined in Algorithm~\ref{alg:efficient_search}. 
This process explores key parameters in $\mathcal{P}$ (e.g., epochs, sample size) that impact both index building and query efficiency (Line 2).
To manage computational costs, an early stopping threshold $T_{build}$ halts the search if the build cost exceeds a set limit. The function \texttt{is\_build\_required()} compares two configurations $\mathbf{p}$ and $\mathcal{P}_{opt}$, returning \texttt{true} if $\mathbf{p}$ could be less expensive for training.
When a configuration $\mathbf{p}$ results in 
a build cost below $T_{build}$, it is then compared with $\mathcal{P}_{opt}$. 
This method guarantees the selection of configurations that satisfy both build and query performance criteria.

\subsection{Implementation of Index Building}
We implement all indices in IBM to ensure consistency and fair comparison. For each RLESI, we reuse its original RL model and adopt the original query algorithms. In contrast, LSIs introduce entirely new index structures that require customized implementations. We thus follow their original implementation in libspatialindex.

\subsubsection{Index Nodes}
In IBM, the index structures of all baselines are categorized 
into two main components: \textit{inner} nodes and \textit{leaf} nodes. Inner nodes, including all non-leaf nodes from the root to the second last layer, do not store spatial objects but act as organizational structures to support search and traversal. 

The maximum node capacity is set to 100 entries (i.e., $B=100$), with each entry occupying 40 bytes: 32 bytes for the coordinates of an MBR and 8 bytes for either a pointer to disk storage (in non-leaf nodes) or a data ID (in leaf nodes). 
The remaining space is allocated to storing metadata, such as the level of the node and the actual number of entries.
The block size is set to 4 KB to ensure that each node fits within a single block, following the default settings from existing works~\cite{rank_space,Packing_Rtrees_SFC}.
For \lisa~\cite{LISA} and \zmindex~\cite{ZM}, we use a large leaf node capacity, i.e., $B=10,000$ to boost space utilization. This avoids the inefficiency of space usage when using $B=100$, which can cause utilization to drop below 10\% due to recursive partitioning after reaching node limits.

\subsubsection{Insertion Strategy}
We implement the insertion strategy according to the original designs of each index.
\rtree~\cite{rtree} is built using an insertion-based strategy, which we follow for \rlrtree~\cite{RLR-tree} and \platon~\cite{PLATON}. \rstar~\cite{r_star_tree} enhances \rtree\ by incorporating a reinsertion technique for node splitting.
For \kdtree~\cite{kdtree}, a point is inserted by performing a point query and adding the point to the target leaf node. If the leaf node is full, it is partitioned using the same procedure as in index building. Since no specific insertion methods are proposed for \qdtree~\cite{qdtree} and \kdgreedy~\cite{qdtree}, we reuse the insertion method of \kdtree~\cite{kdtree}. 
For MP-based indices, maintaining order during insertion is challenging due to their reliance on sorted spatial objects during index building. As a result, we reuse the insertion method of \rtree~\cite{rtree} for these indices, given their \rtree-based structure.

\subsection{Selection of Baselines}\label{subsec:baselines}

\begin{figure}[h]
  \centering
  \includegraphics[width=0.50\textwidth]{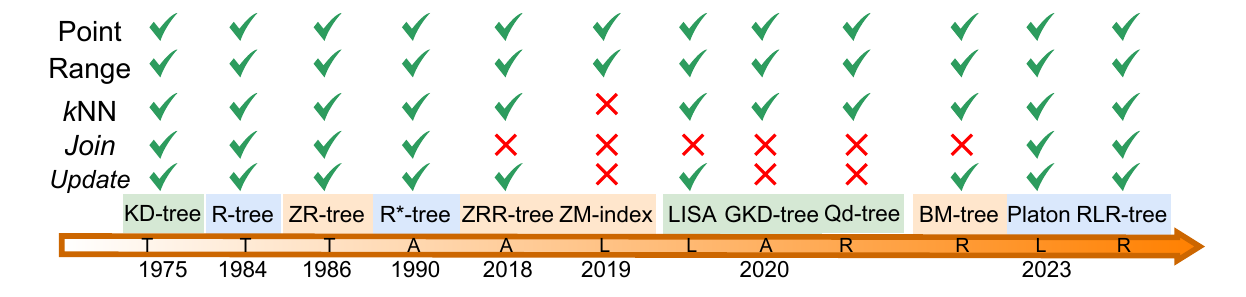}
  \caption{Chronological listing of selected indices.}
  \label{fig:history}
\end{figure}

The vast number of spatial indices makes comparing them all unfeasible. Thus, we prioritize representative indices. 
Figure~\ref{fig:history} summarizes the selected indices in chronological order and groups them by category using background colors: DP-based (blue), SP-based (green), and MP-based (orange). 
Beneath each index, we tag its type with T/A/L/R for \underline{T}raditional / \underline{A}dvanced / \underline{L}earned / \underline{R}L-enhanced. For capabilities, we indicate native support for point, range, $k$NN, spatial join, and updates using \cmark\ and \xmark.

We ensure fairness and representativeness by selecting indices across all categories.  
For RLESIs, we choose one state-of-the-art representative per family: \rlrtree~\cite{RLR-tree} (DP, only open-sourced option), \qdtree~\cite{qdtree} (SP, only RL-enhanced variant), and \bmtree~\cite{BMTree} (MP, LMSFC~\cite{LMSFC} lacks open-source).  
Each RLESI is compared against its direct traditional counterpart: \rtree~\cite{rtree}, \kdtree~\cite{kdtree}, and \zrtree~\cite{ZRtree}.  
Advanced baselines extend these families: \rstar~\cite{r_star_tree} for \rtree, \kdgreedy~\cite{qdtree} for \kdtree, and \zrrtree~\cite{rank_space} for \zrtree.  
For LSIs, we select strong representatives: \platon~\cite{PLATON} (DP), \lisa~\cite{LISA} (SP), and \zmindex~\cite{ZM} (MP).  
This setup ensures coverage of traditional, advanced, learned, and RL-enhanced methods across DP-, SP-, and MP-based categories.

\section{Experimental Setup}\label{sec:exp_setup}
All experiments are executed on a server 
running 64-bit Ubuntu 20.04 with a 3.60 GHz Intel i9 CPU, 64~GB RAM, and a 1 TB HDD. 
We use \emph{PyTorch} 1.4~\cite{pytorch} to train models for RLESIs based on the CPU
and employ its C++ APIs to load models and invoke index building. 

\vspace{-0.5em}
\subsection{Datasets}
\begin{figure}[h]
  \centering
\includegraphics[width=0.45\textwidth]{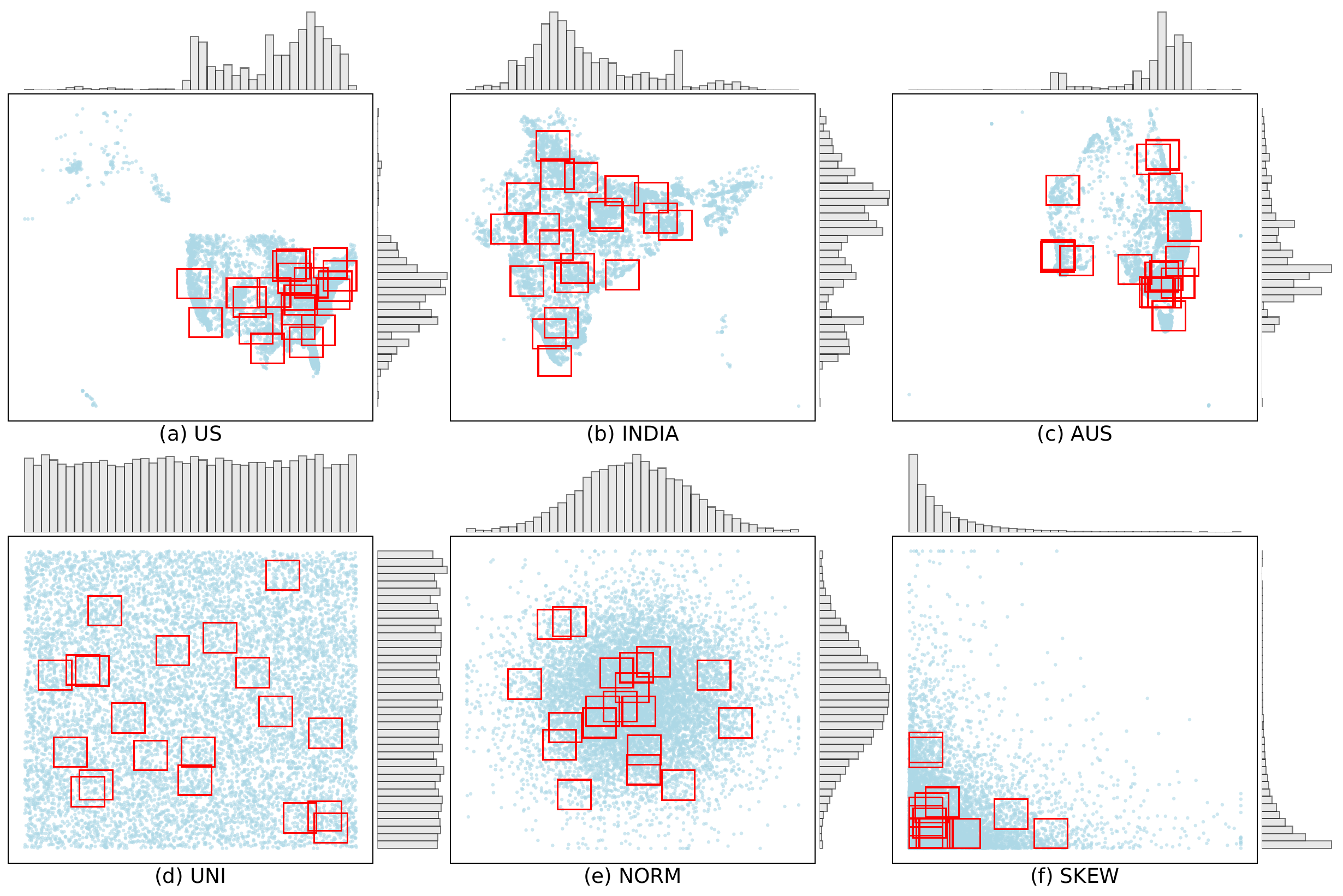}
  \caption{Six datasets used in the experimental study.} 
  \label{fig:all_dataset_10000_hist_range}
\end{figure}

We evaluate all baselines on six 2D datasets: three synthetic datasets and three real-world datasets derived from OpenStreetMap (OSM)~\cite{osm_stats}, each containing 100M spatial points following configurations commonly used in prior work~\cite{rank_space, RSMI, LISA, RLR-tree}. 
Figure~\ref{fig:all_dataset_10000_hist_range} illustrates the spatial distributions by sampling 10,000 points (blue) and overlaying 20 example range queries (red), which are generated by randomly sampling 20 centers following the data distribution and wrapping them with axis-aligned rectangles.
Histograms on the $x$ and $y$ axes further highlight the distinct distributional characteristics.

The three synthetic datasets represent uniform, normal, and skewed distributions, which are widely used in recent studies~\cite{PLATON,RLR-tree,BMTree,LBMC,rank_space,LISA}. 
For example,
\platon, \rlrtree, and \zrrtree\ evaluate across all three distributions, whereas \lisa\ is limited to skewed and uniform datasets, and \bmtree\ reports results on normal and uniform distributions.
We use 100M points for all to ensure consistency with the scale of our real datasets.

For real datasets,
we select three OSM datasets (US, India, and Australia) to capture diverse geographic patterns. 
The US and India datasets have been widely used, for example, by \platon\ (US and India), \rlrtree\ (India), and \bmtree\ (US). 
Here, the US dataset reflects dense metropolitan clusters with sparse rural coverage, the India dataset exhibits a more balanced spread across cities and towns. 
The Australia dataset has not been used in prior work, but it provides a skewed distribution with points concentrated along coastal cities. 

\vspace{-0.5em}
\subsection{Workloads}
We test all the indices on seven distinct workload types:
\begin{enumerate}[leftmargin=6mm]
\item \textbf{Point Query-Only:} Executes 200,000 point queries sampled from the corresponding dataset, following~\cite{sosd}.
\item \textbf{Range Query-Only:} Executes 1,000 range queries with centroids sampled like point queries. The default query shape is square, with edge lengths varying as \{0.01\%, 0.05\%, \textbf{0.1\%}, 0.5\%, 1\%\} of the data space~\cite{BMTree}. Additionally, aspect ratios vary across \{1/16, 1/4, \textbf{1}, 4, 16\}, with \{1/4, 1, 4\} matching settings in \bmtree~\cite{BMTree}, and {1/16, 16} added to explore skewed queries.
\item \textbf{$k$NN Query-Only:} Executes 1,000 $k$NN queries with query points sampled similarly to point queries. Values of $k$ are chosen from \{1, 5, \textbf{25}, 125, 625\}, following configurations in~\cite{RLR-tree}.
\item \textbf{Spatial Join-Only:} 
Executes 1,000 spatial joins using the classical hierarchical join algorithm~\cite{R_tree_join}. 
We use range-based predicates, as they naturally define the spatial overlap conditions required for joins.
\item \textbf{Write-Only:} Inserts 10M randomly generated data points after the initial bulk-loading of 10M points.
\item \textbf{Write-Heavy:} Performs 90\% inserts and 10\% lookups after bulk-loading 10M random data points. For every 20 operations, 18 are inserts and 2 are lookups.
\item \textbf{Read-Heavy:} Performs 90\% lookups and 10\% inserts after bulk-loading 10M random data points. For every 20 operations, 2 are inserts and 18 are lookups.
\end{enumerate}

Write-only, write-heavy, and read-heavy workloads follow the configurations in~\cite{ULI_Evaluation,LI_disk}, adapted for spatial data. 
In our experiments, the query range with a 0.1\% edge length and an aspect ratio of 1 is the default setting for building RLESIs, \kdgreedy, and \platon. Default parameters across workloads are highlighted in \textbf{bold}.

\vspace{-0.5em}
\subsection{Metrics}
We use the following metrics to evaluate index performance:
\begin{enumerate}[leftmargin=6mm]
\item \textbf{Index Building Cost:} Measures the total time required to build the index from reading the dataset to storing it on disk. For RLESIs, this includes model training.
\item \textbf{Average I/O per Search Query:} Captures the average number of disk blocks accessed for each query. This is divided into \textbf{leaf I/O}, representing data storage nodes, and \textbf{inner I/O}, which accounts for index nodes. This metric provides a detailed breakdown of I/O efficiency.
\item \textbf{Average Query/Insert Latency:} Reports the average response time for querying or inserting data into the index.
\item \textbf{Latency Distribution (P1 to P99):} Tracks latency percentiles from the 1st to the 99th, including \textbf{tail latency} (P99). This helps evaluate both typical and worst-case performance, as RLESIs may exhibit suboptimal behavior outside trained conditions.
\item \textbf{Index Statistics:} Includes tree height, index size, node utilization, and metrics specific to insertion, such as the number of node splits or adjustments. 
These statistics help explain performance outcomes, e.g., how frequent splits impact insertion latency.
\end{enumerate}

\section{Experimental Results}\label{sec:exp_results}
To evaluate the effectiveness of RLESIs, we conduct a series of experiments that aim to answer the following key questions:
\noindent\textbf{Q1:} How do RLESIs compare with traditional, advanced, and learned indices in terms of query performance?  
(Section~\ref{subsec:query_performance} and Section~\ref{subsec:query_variation})
\noindent\textbf{Q2:} How do all indices perform under worst-case scenarios, such as tail latency?
(Section~\ref{subsec:tail_latency}) 
\noindent\textbf{Q3:} How well do the indices handle insertions, especially under write-heavy workloads?  
(Section~\ref{subsec:insertion}) 
\noindent\textbf{Q4:} How do key index statistics (e.g., size, height, splits) evolve during index building?  
(Section~\ref{subsec:index_build})
\noindent\textbf{Q5:} How does dataset cardinality affect performance across different index types?  
(Section~\ref{subsec:cardinality_variation})
\noindent\textbf{Q6:} To what extent can parameter tuning improve RLESI performance? 
(Section~\ref{subsec:rlesi_tuning})

For better visualization, indices in three categories are colored blue (DP-based), orange (MP-based), and green (SP-based). 
Within each group, four types (traditional, advanced, learned, and RL-enhanced) are filled by different patterns, and the same type uses consistent patterns across groups.

\subsection{Query Performance Study}\label{subsec:query_performance}
We present latency and I/O results for range queries. For other query types, we report latency only, as I/O trends closely align with latency, allowing for brevity.
All RLESIs have been tuned, with details outlined in Section~\ref{subsec:rlesi_tuning}.

\begin{figure}[h]
  \centering
\subfloat[Query latency~\label{fig:range_query_time}]   {
\includegraphics[width=0.45\textwidth]{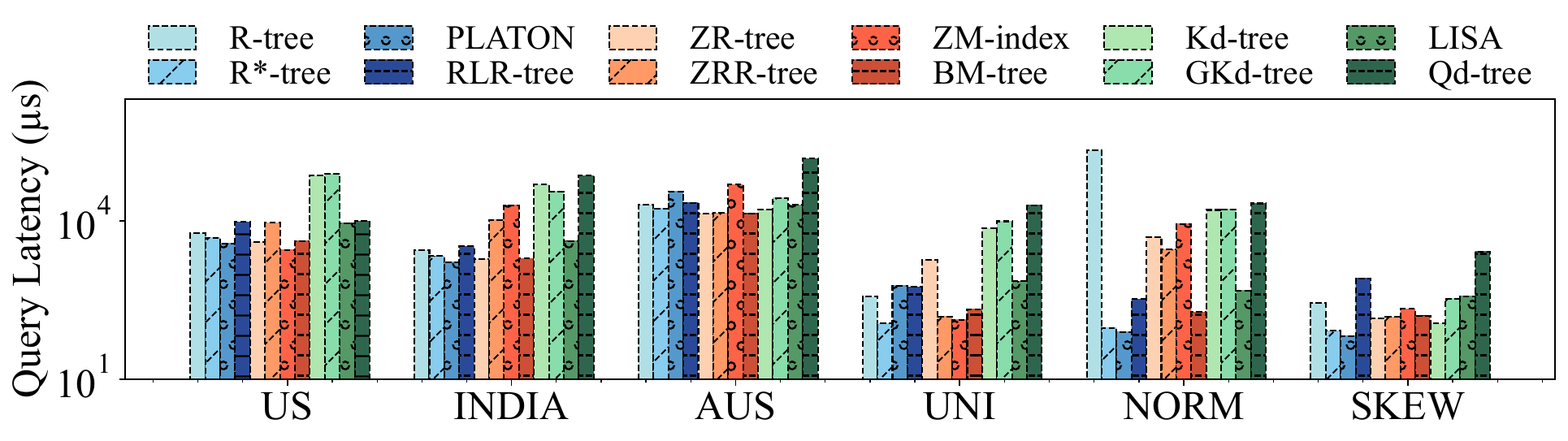}
}\\
\vspace{-1em}
\subfloat[Query I/O~\label{fig:range_query_IO}]   {
\includegraphics[width=0.45\textwidth]{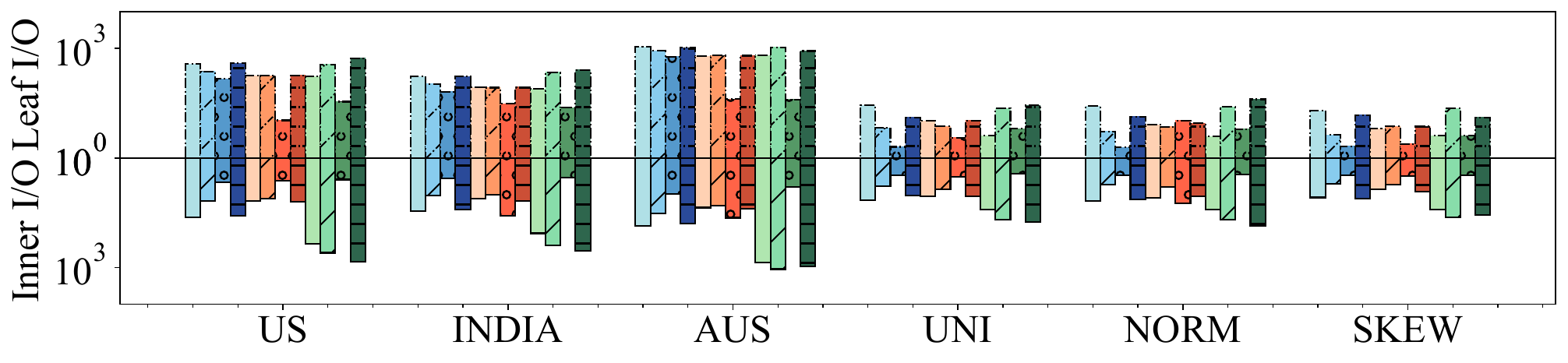}
}
  \caption{Range query} 
  \label{fig:range_query}
\end{figure}

\subsubsection{Range Query-Only}
Figure~\ref{fig:range_query} presents the range query latency and I/O count across six datasets. We observe that:

\stepcounter{observation} \noindent\textbf{O\theobservation: Across all datasets, RLESIs show competitive query performance compared to traditional indices but fall short of advanced and learned baselines.}
\rlrtree\ achieves comparable performance to \rtree\ in terms of query latency and I/O. However, \rlrtree\ does not surpass \rstar\ or \platon\ in range query performance. This limitation arises because \rlrtree\ relies on \rtree\ for node splits, leading to suboptimal overlap management compared to \rstar\ and \platon. As shown in Figure~\ref{fig:range_query_IO}, \rstar\ and \platon\ achieve fewer leaf and inner I/Os than \rtree\ and \rlrtree.

\bmtree\ performs similarly to \zrtree\ across real datasets, with much better query performance on synthetic datasets. This advantage is attributed to \bmtree\ learning from the mapped values of spatial data points, which can enhance the quality of the learned model. Compared to \zrrtree\ and \zmindex, \bmtree\ exhibits comparable query latency and I/O on most datasets. However, \bmtree\ outperforms others in query latency on the NORM dataset, while the I/O cost remains relatively close. This suggests that NORM (see Figure 6e) can cause structural traversal differences for \zrtree\ and \zrrtree.

\qdtree\ and \kdgreedy\ show inferior query latency and I/O compared to \kdtree\ and \lisa\ across most datasets. 
This is attributed to the RL model in \qdtree\ failing to converge effectively, as indicated by training logs, which is also a challenge highlighted in~\cite{RLR-tree}.
For \kdgreedy, the partitioning strategy
heavily relies on query boundaries due to its greedy nature, which limits the generalizability.

\stepcounter{observation}
\noindent\textbf{O\theobservation:
RLESIs show consistent alignment between query latency and I/O across datasets as well as other indices except for \zmindex\ and \lisa.
}
\zmindex\ and \lisa\ show higher latency but lower I/O on all datasets due to the impact of skewed data on model prediction accuracy, which causes more computational costs. For \zmindex\ and \lisa, inner I/O is solely associated with recursive model predictions rather than tree traversal, leading to no overlap between these operations. Additionally, \zmindex\ and \lisa\ benefit from large leaf node capacity, which reduces the number of leaf I/Os required.

\begin{figure}[h]
  \centering
  \includegraphics[width=0.45\textwidth]{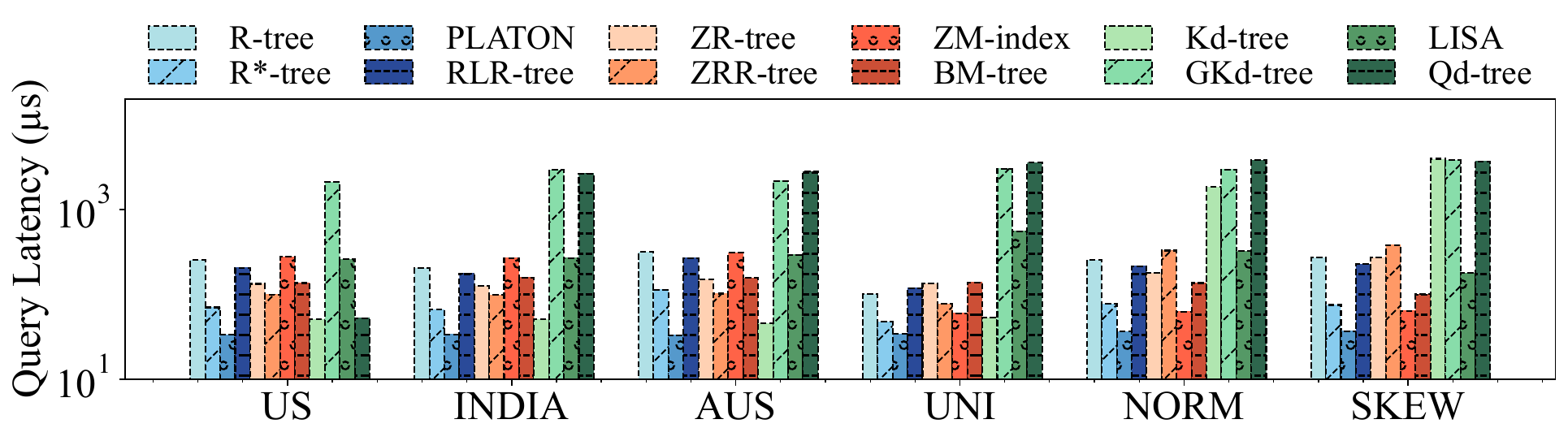}
  \caption{Point query latency}
  \label{fig:point_query}
\end{figure}

\subsubsection{Point Query-Only}

Figure~\ref{fig:point_query} illustrates the point query latency across six datasets. We observe that:

\stepcounter{observation}
\noindent\textbf{O\theobservation: Advanced and learned indices consistently outperform RLESIs in query latency across all datasets.}
RLESIs offer moderate improvements over traditional indices in specific cases. 
\rlrtree\ achieves slightly better query latencies than \rtree, while \bmtree\ shows competitive performance relative to \zrtree, though not always superior.
There is a special case where on the US dataset, \qdtree\ shows low query latency, this is due to the tuning, which is also shown in Figure~\ref{fig:all_query_time_build_time}.
In comparison, 
advanced indices, such as \rstar\ and \zrrtree, benefit from optimized structures, delivering balanced performance across various query types, including point queries. Among learned indices, \platon\ achieves the best query latency. \zmindex\ performs exceptionally well on synthetic datasets due to their predictable patterns but incurs higher latency on real-world datasets, where data is more challenging to model~\cite{ZM,RSMI}.

\begin{figure}[h]
  \centering
  \includegraphics[width=0.45\textwidth]{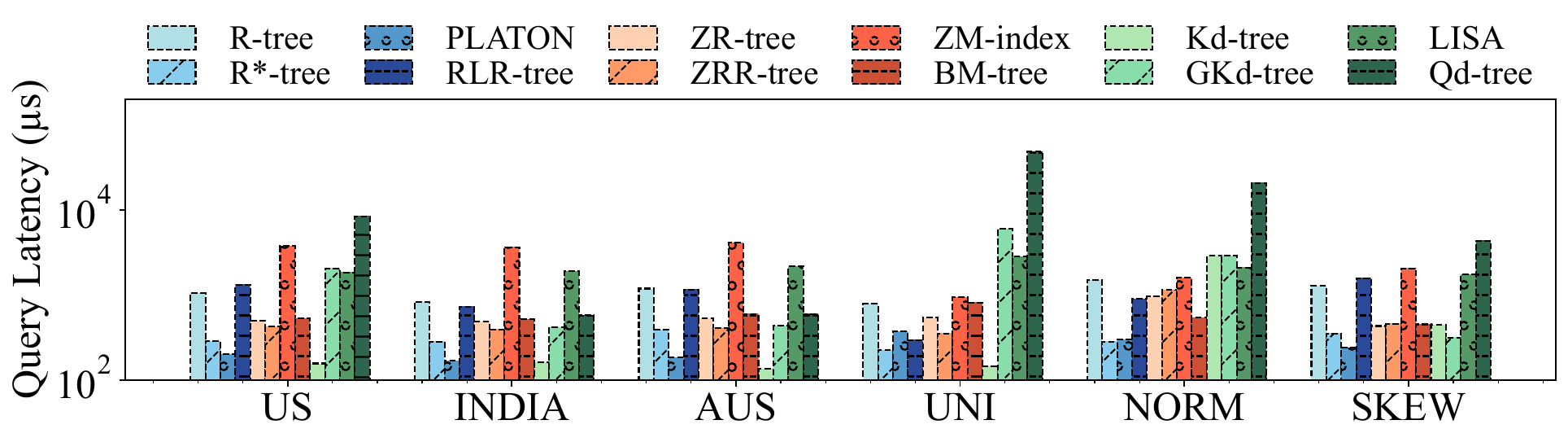}
  \caption{$k$NN query latency}
  \label{fig:knn_query_time}
\end{figure}

\subsubsection{$k$NN Query-Only}

Figure~\ref{fig:knn_query_time} shows the $k$NN query latency across different datasets. We observe that:

\stepcounter{observation}
\noindent\textbf{O\theobservation:
Among RLESIs, \bmtree\ achieves the best $k$NN query latency across most datasets; however, \kdtree\ and \platon\ still outperform all RLESIs on most datasets.}
RLESIs fail to deliver competitive $k$NN query latency, particularly \rlrtree\ and \qdtree, which do not effectively partition the data space. However, \bmtree\ achieves better performance by clustering closely located data points and packing them consecutively into blocks, optimizing query latency.
Overall, RLESIs lag behind advanced and learned baselines in $k$NN query latency, which is consistent with observations in range queries.
The superior performance of \kdtree\ and \platon\ is attributed to minimal or no node overlap, reducing the need to visit unnecessary nodes during $k$NN queries. \zmindex\ underperforms for $k$NN queries due to the absence of a dedicated $k$NN algorithm, as it reuses the method from \rtree.

\begin{figure}[h]
  \centering
  \includegraphics[width=0.45\textwidth]{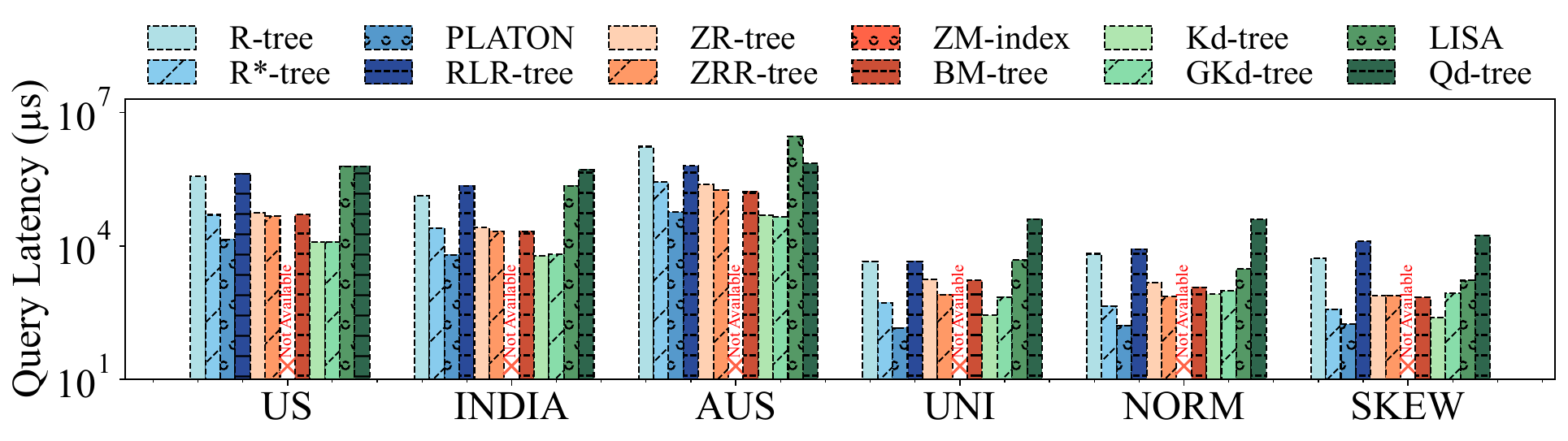}
  \caption{Spatial join latency}
  \label{fig:join_query_time}
\end{figure}

\subsubsection{Spatial Join-Only}
Figure~\ref{fig:join_query_time} illustrates the spatial join 
latency across six datasets. We observe that:

\stepcounter{observation}
\noindent\textbf{O\theobservation:
RLESIs are competitive with traditional indices and often, though not always, behind advanced and learned indices.}
\rlrtree\ can only outperform R-tree on AUS dataset, while being slower on the other datasets.
This stems from \rlrtree’s suboptimal data partitioning, as illustrated also in Figure~\ref{fig:range_query_time}.
\bmtree\ performs comparably to \zrtree\ and \zrrtree, as \bmtree\ minimizes overlap and backtracking, yielding fewer overlaps.
\qdtree\ underperforms its counterparts as it derives partitions from a subset of queries, which can only learn substantial partitions.

\begin{figure}[h]
  \centering
\subfloat[Varying edge length of queries~\label{fig:range_query_time_varying_range}]   {
\includegraphics[width=0.45\textwidth]{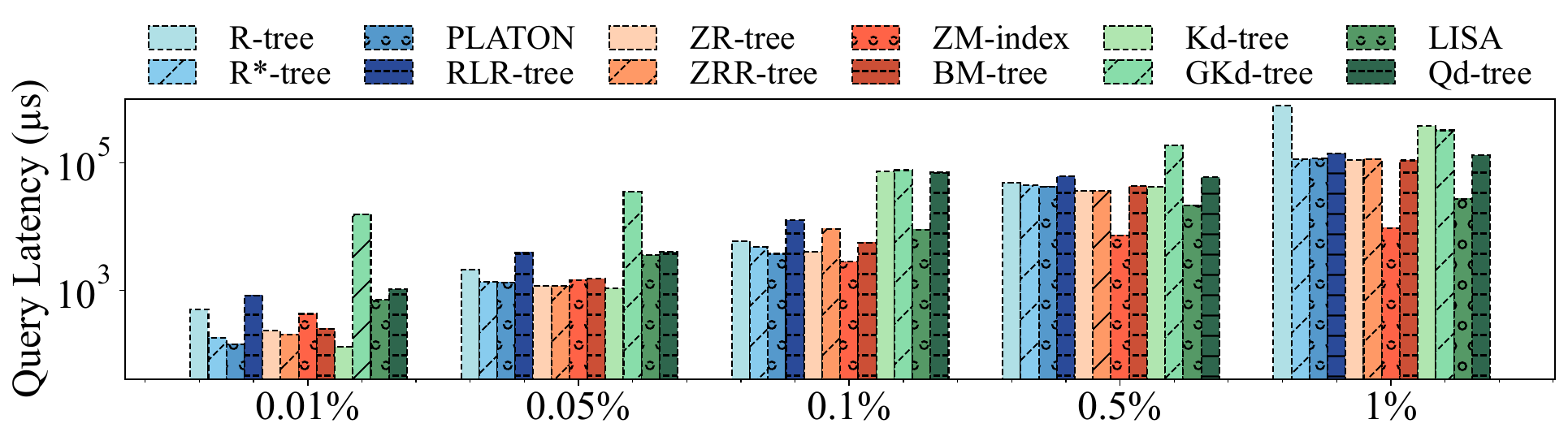}
}\\
\vspace{-1em}
\subfloat[Varying aspect ratio of queries~\label{fig:range_query_time_varying_aspect_ratio}]   {
\includegraphics[width=0.45\textwidth]{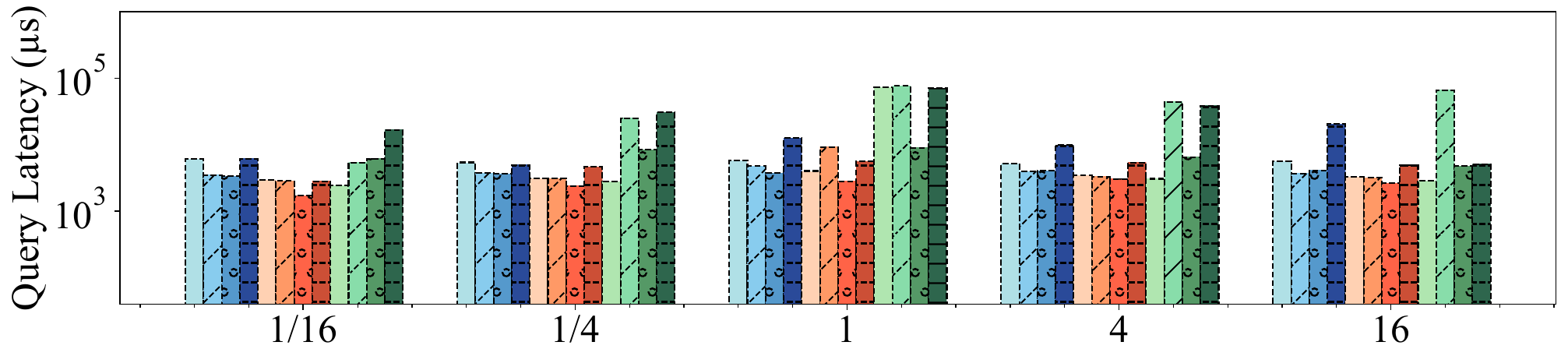}
}\\
\vspace{-1em}
\subfloat[Varying $k$~\label{fig:knn_query_time_varying_k}]   {
\includegraphics[width=0.45\textwidth]{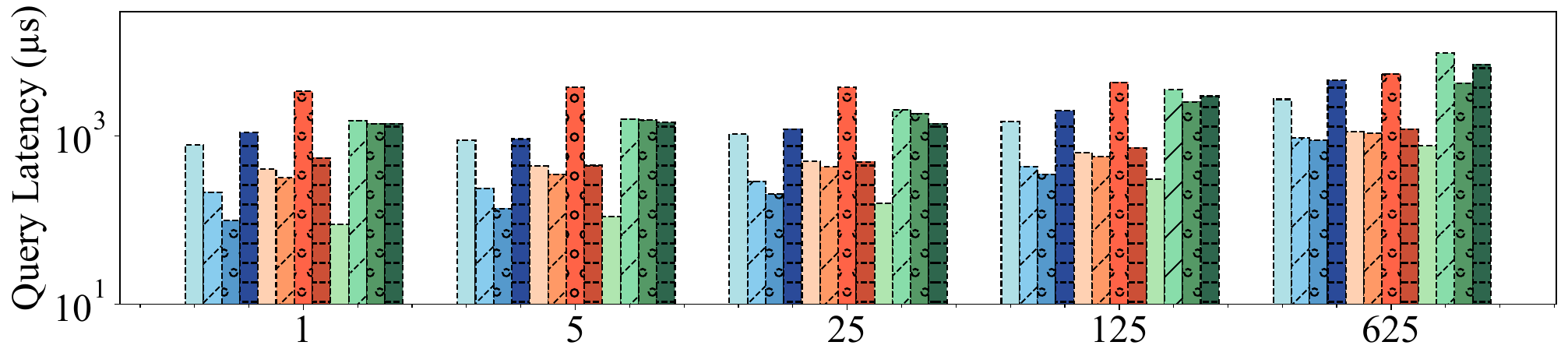}
}
\caption{Varying queries} 
\label{fig:varying_query}
\end{figure}


\subsection{Query Variation Study}\label{subsec:query_variation}

\subsubsection{Varying Query Range Size}

We execute range queries with varying edge lengths on the US dataset, with query latency results shown in Figure~\ref{fig:range_query_time_varying_range}. We observe that:

\stepcounter{observation}
\noindent\textbf{O\theobservation: RLESIs and \kdgreedy\ exhibit higher range query latency for small query ranges.}
As the query range increases, query latency rises for all indices due to the larger number of data points involved. RLESIs show inferior performance when the edge length is 0.1\% or smaller, particularly for \rlrtree\ and \qdtree. However, as the edge length grows, RLESIs demonstrate moderate latency improvements.
In comparison, \kdgreedy, an advanced baseline, exhibits worse latency across all ranges. Among LSIs, \platon\ delivers consistently competitive latency across all edge lengths, while \zmindex\ shows outstanding performance as the edge length exceeds 0.1\%.

\subsubsection{Varying Query Range Aspect Ratio}
We execute range queries with varying aspect ratios on the US dataset, presenting the query latency in Figure~\ref{fig:range_query_time_varying_aspect_ratio}. We observe that:

\newcounter{previousObservation}
\setcounter{previousObservation}{\value{observation}}
\stepcounter{observation}
\noindent\textbf{O\theobservation: 
RLESIs and \kdgreedy\ are sensitive to query aspect ratios.} RLESIs and \kdgreedy\ continue to exhibit inferior query latency when the aspect ratio of query ranges varies. \rlrtree\ outperforms \rtree\ when the aspect ratio is 1/4 and 1/16, but still falls short of \rstar\ and \platon. Notably, the query latency of \rlrtree\ increases significantly when the aspect ratio is 16. Similarly, \bmtree\ falls behind other MP-based counterparts, due to its quadtree partitioning, which is less effective for skewed queries.

\subsubsection{Varying $k$}
We execute $k$NN queries on the US dataset with 
$k$ varying from 1 to 625 in Figure~\ref{fig:knn_query_time_varying_k}.
We observe that:

\stepcounter{observation}
\noindent\textbf{O\theobservation: 
Except for \bmtree, RLESIs exhibit the highest latency increase with larger $k$ values, while \kdtree\ and \platon\ remain efficient.
}
As $k$ increases, the query latency of all indices rises, and the result pattern does not change much. For RLESIs, \rlrtree\ and \qdtree\ show the most pronounced increase in latency, indicating their inefficiency in handling large $k$ values. This inefficiency arises because 
\rlrtree\ requires more computations
and
\qdtree\ does not partition the data space effectively, requiring more leaf node accesses as $k$ increases. 
In contrast, \kdtree\ and \platon\ are efficient in handling $k$NN queries, while another LSI, \zmindex, 
shows inferior performance due to its missing implementation of $k$NN query method and the reuse of $k$NN query from \rtree.

\begin{figure}[h]
  \centering
\subfloat[Range query~\label{fig:range_query_P99}]   {
\includegraphics[width=0.45\textwidth]{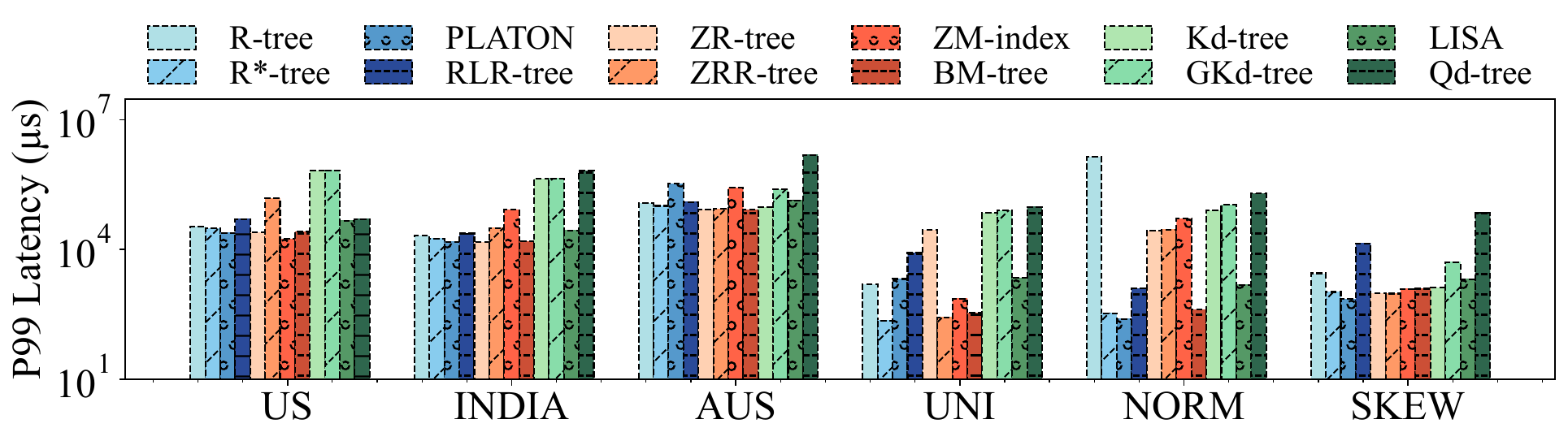}
}\\
\vspace{-1em}
\subfloat[Point query~\label{fig:point_query_P99}]   {
\includegraphics[width=0.45\textwidth]{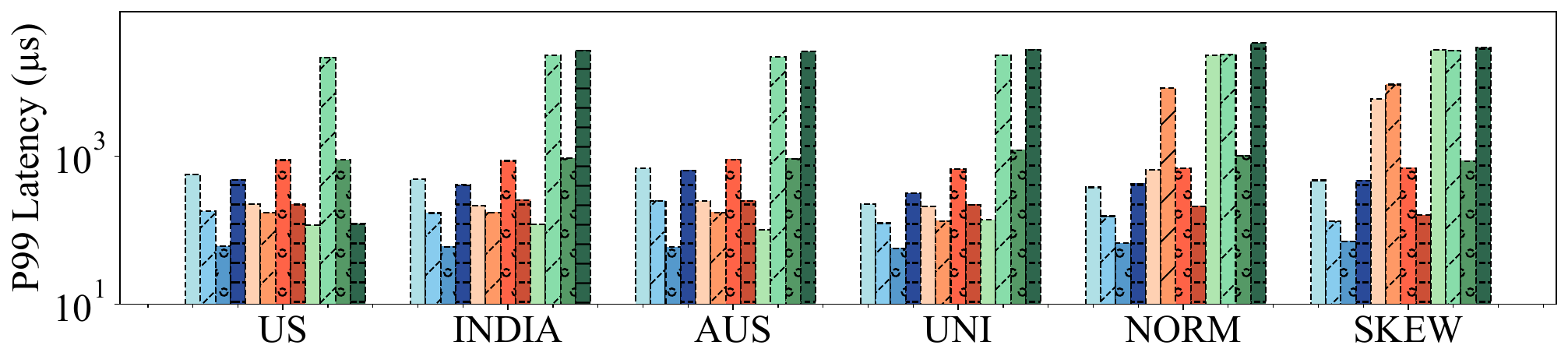}
}\\
\vspace{-1em}
\subfloat[$k$NN query~\label{fig:knn_query_P99}]   {
\includegraphics[width=0.45\textwidth]{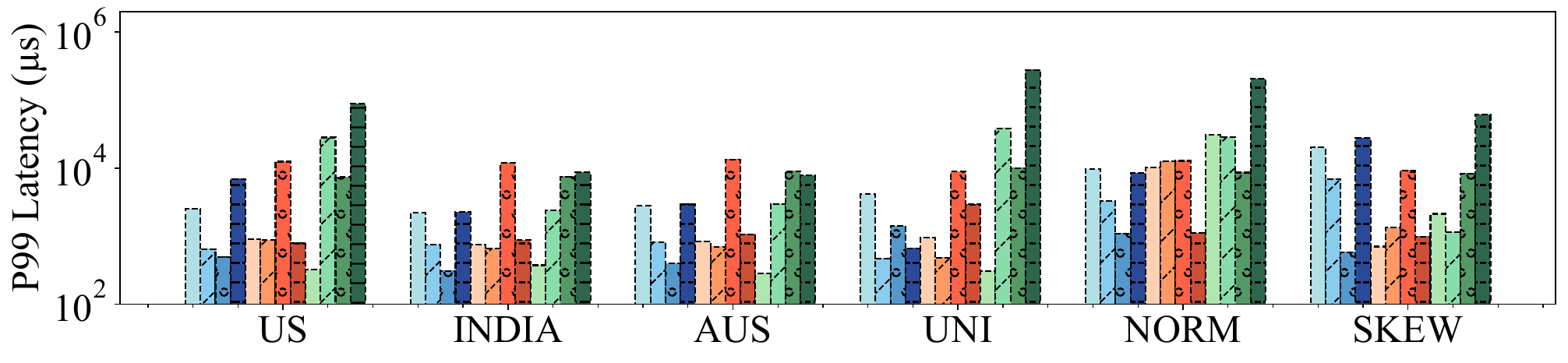}
}
  \caption{P99 Query latency} 
  \label{fig:P99_latency}
\end{figure}

\subsection{Latency Study}\label{subsec:tail_latency}

\subsubsection{Tail Latency}
Figure~\ref{fig:P99_latency} presents P99 latency for point, range, and $k$NN queries across all datasets. We observe that:

\stepcounter{observation}
\noindent\textbf{O\theobservation:
RLESIs show relatively high P99 latency, especially \rlrtree\ and \qdtree, while traditional indices maintain moderate.}
This variability comes from the RLESI building. For instance, \bmtree\ performs relatively well as its MP-based structure balances the tree effectively and mitigates the impact of worst-case queries. Conversely, \rlrtree\ and \qdtree\ often exhibit poor performance, with \rlrtree\ struggling in most scenarios and \qdtree\ performing significantly worse.
In contrast, traditional indices like \kdtree\ and \rstar\ demonstrate more consistent P99 latency. This is attributed to their robust design, e.g., \kdtree\ uses recursive median splitting, and \rstar\ employs reinsertion strategies to handle overflows. These methods ensure effective partitioning of data and space, minimizing imbalances and maintaining stable query.

\begin{figure}[h]
  \centering
\subfloat[Range query~\label{fig:range_query_percentiles}]   {
\includegraphics[width=0.45\textwidth]{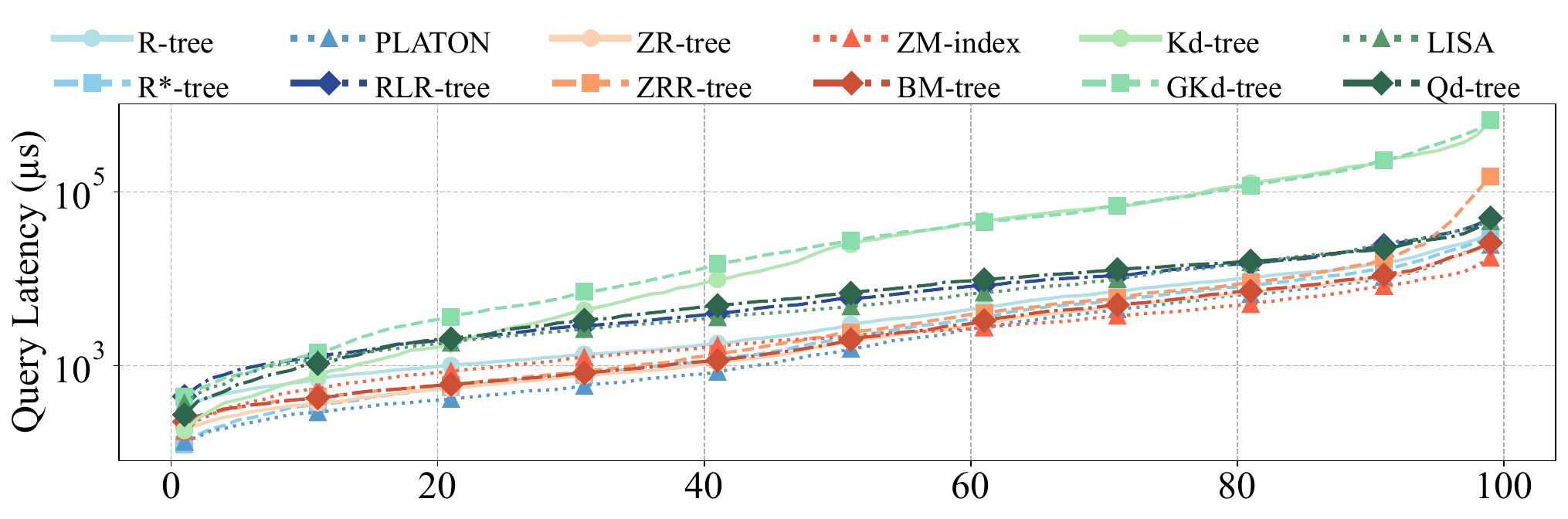}
}\\
\vspace{-1em}
\subfloat[Point query~\label{fig:point_query_percentiles}]   {
\includegraphics[width=0.45\textwidth]{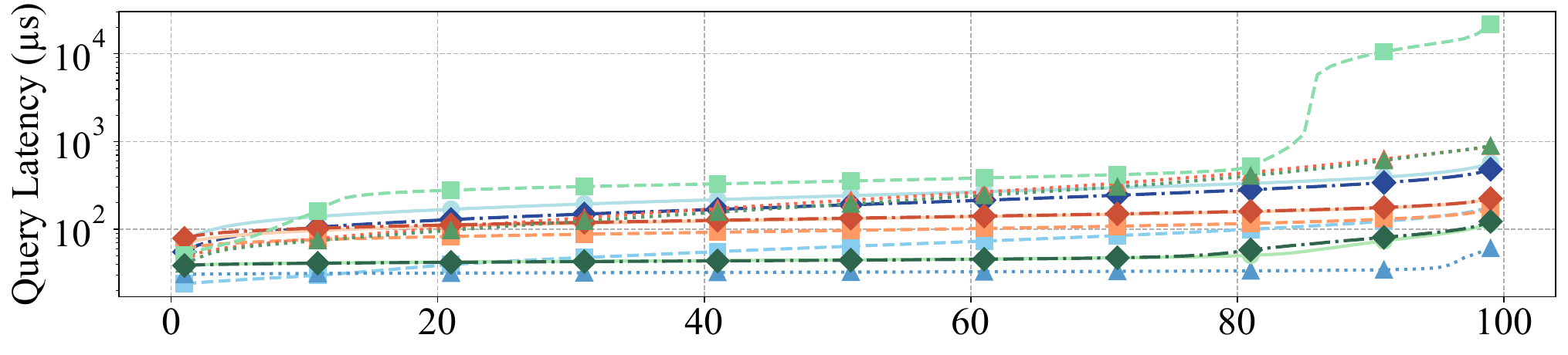}
}\\
\vspace{-1em}
\subfloat[$k$NN query~\label{fig:knn_query_percentiles}]   {
\includegraphics[width=0.45\textwidth]{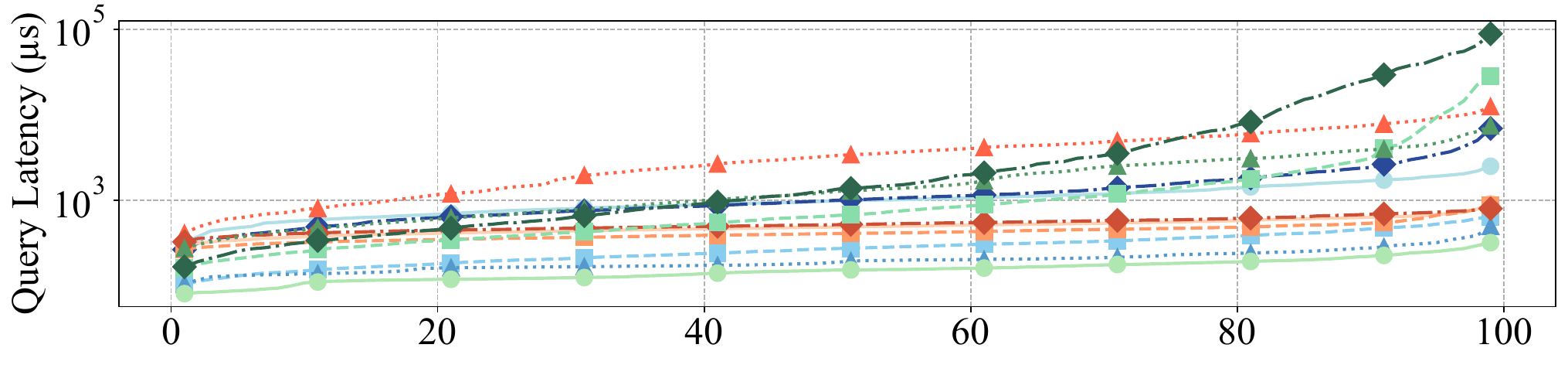}
}
  \caption{Query latency distribution from P1 to P99} 
  \label{fig:P1-99_latency}
\end{figure}

\subsubsection{Latency Distribution}
Figure~\ref{fig:P1-99_latency} shows the P1 to P99 latency distribution on the US dataset. We observe that:

\stepcounter{observation}
\noindent\textbf{O\theobservation: RLESIs exhibit consistent range and point query latency increases from P1 to P99, comparable to traditional indices.}
Query latency varies due to differences in data density and node overlap across spatial regions. Denser areas lead to more node accesses and higher latency, especially when queries intersect overlapping partitions. \rlrtree\ and \bmtree exhibit similar latency distributions to traditional indices, demonstrating robust performance across spatial variations.

\stepcounter{observation}
\noindent\textbf{O\theobservation: RLESIs based on DP or SP structures suffer from latency increases at high percentiles for point and $k$NN queries.}
This is caused by query regions intersecting dense clusters or heavily overlapping nodes, which significantly increases the number of accessed nodes. For instance, \rlrtree\ and \qdtree\ exhibit steep latency growth from P90 to P99, similar to traditional DP- and SP-based indices such as \kdgreedy. In contrast, MP-based RLESIs like \bmtree\ maintain smoother latency curves, as their one-dimensional mappings reduce sensitivity to spatial skew.

\begin{figure}[h]
  \centering
\subfloat[Write-Only~\label{fig:write_only}]   {
\includegraphics[width=0.45\textwidth]
{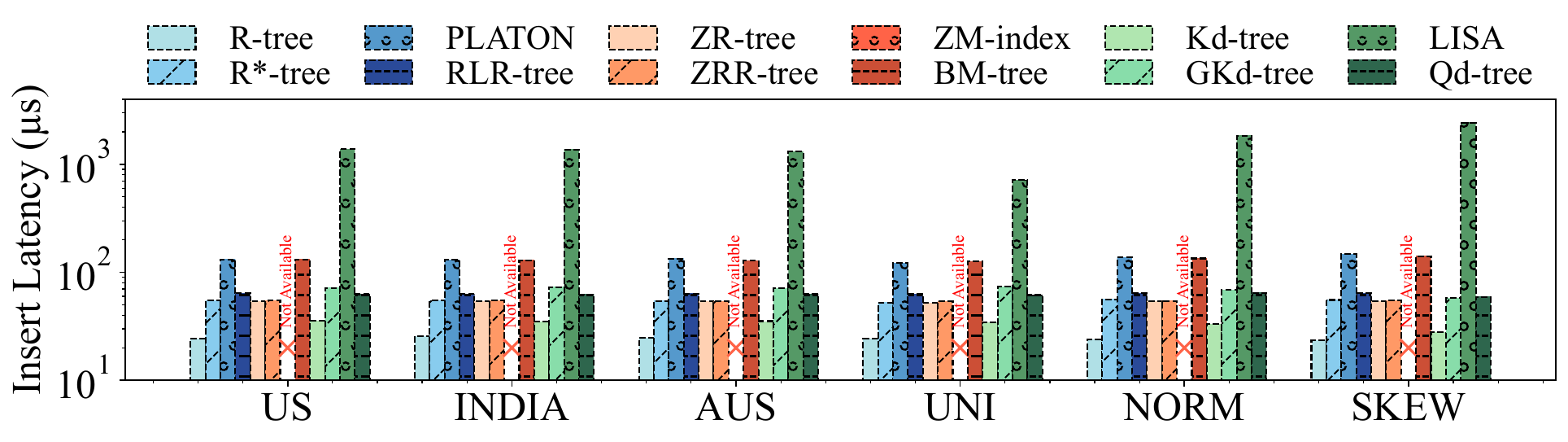}
}\\
\vspace{-1em}
\subfloat[Write-Heavy~\label{fig:write_heavy_insert_time}]   {
\includegraphics[width=0.45\textwidth]{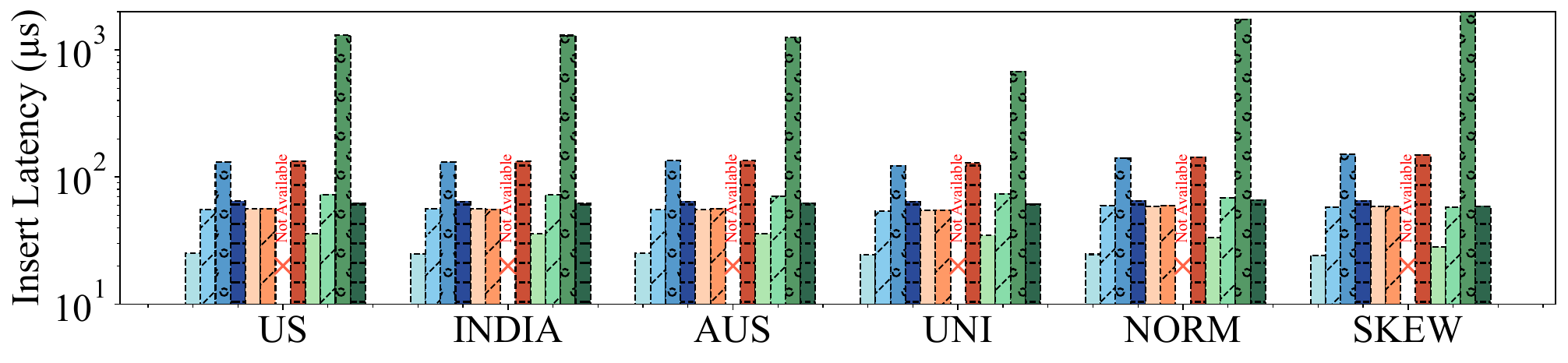}
}\\
\vspace{-1em}
\subfloat[Read-Heavy~\label{fig:read_heavy_insert_time}]   {
\includegraphics[width=0.45\textwidth]{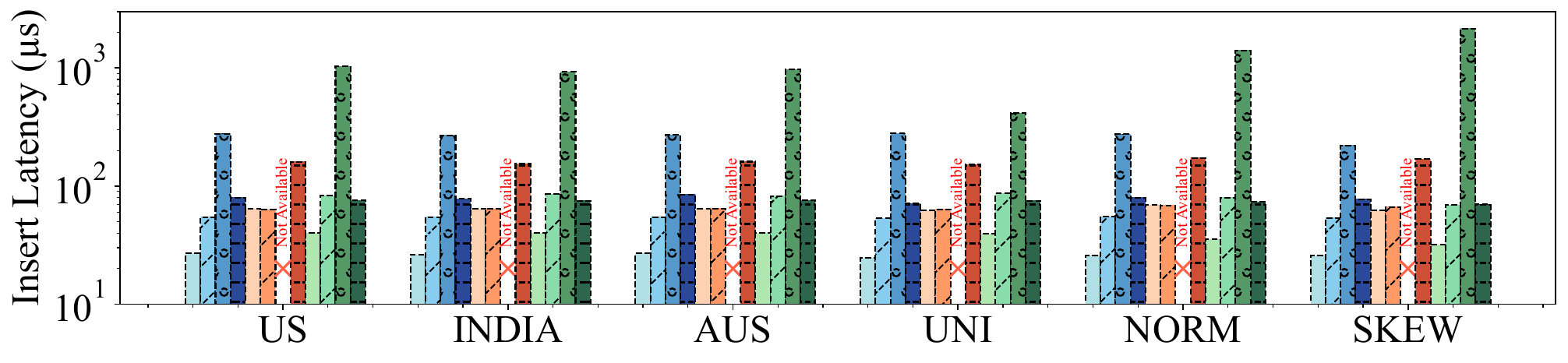}
}
  \caption{Insertion latency for different workload types} 
  \label{fig:insertion_latency}
\end{figure}

\subsection{Insertion Performance Study}\label{subsec:insertion}
\subsubsection{Insertion Latency}
Figure~\ref{fig:insertion_latency} shows insertion latency for ready-heavy, write-heavy, and write-only workloads.
We observe that:

\stepcounter{observation}

\noindent\textbf{O\theobservation: All RLESIs and LSIs exhibit high insertion latency, with \lisa\ performing the worst.}
Insertions trigger node splits and new node creation.
For RLESIs, which add extra costs due to model predictions 
for \rlrtree\ and \qdtree. While \bmtree\ does not rely on model predictions, its densely packed leaf nodes (fill ratio 1.0) necessitate frequent node splits during insertions. In contrast, DP- and MP-based indices reserve more space in their leaf nodes, resulting in fewer node splits. 
Notably, \lisa\ exhibits the highest latency as it creates a model node to handle overflow and reassign points to different leaf nodes.
For \platon, it exhibits the second-highest latency due to its top-down partitioning strategy, which fully packs nodes and intensifies the cost of insertions.

\begin{figure}[h]
  \centering
\subfloat[Write-Only~\label{fig:write_only_splits}]   {
\includegraphics[width=0.45\textwidth]{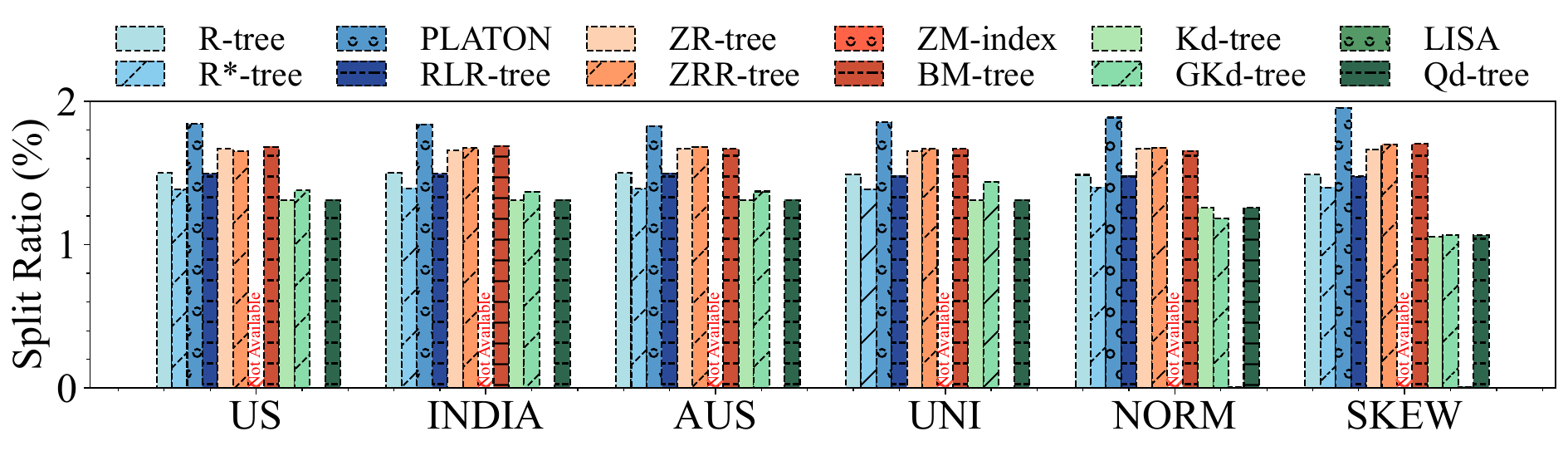}
}\\
\vspace{-3mm}
\subfloat[Write-Heavy~\label{fig:write_heavy_splits}]   {
\includegraphics[width=0.45\textwidth]{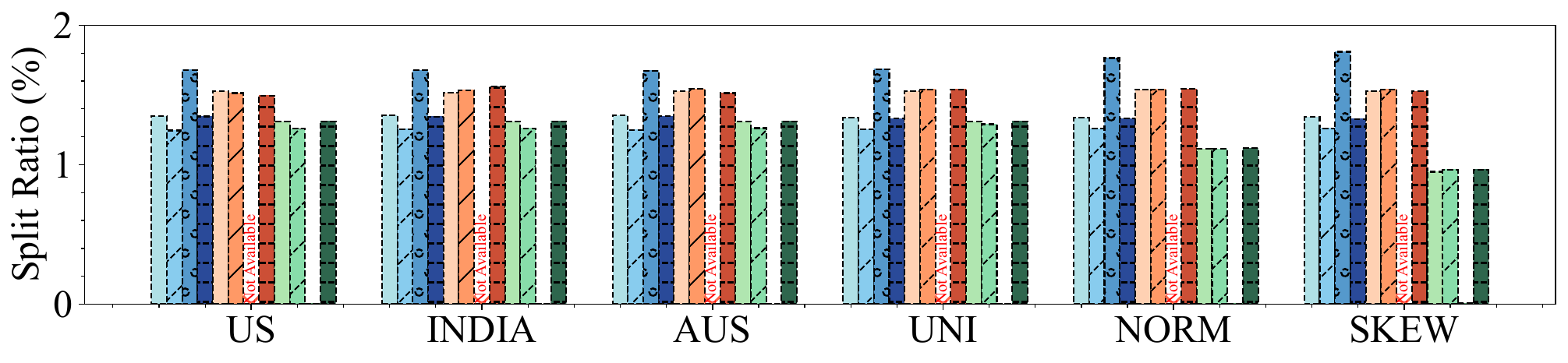}
}
\caption{Node split ratio} 
\label{fig:node_splits_ratio}
\end{figure}

\subsubsection{Node Split Ratio}
We show the node split ratio of write-only and write-heavy workloads in
Figure~\ref{fig:node_splits_ratio}. We observe that:

\stepcounter{observation}
\noindent\textbf{O\theobservation: \platon\ and MP-based indices exhibit the highest split ratio, while SP-based indices have the lowest, especially \lisa.}
As explained for insertion latency, \platon\ and MP-based indices fully utilize their leaf nodes, leaving no unused space. Consequently, inserting new points into these indices often requires a leaf node split with high probability. In contrast, SP-based indices and some DP-based indices retain unused space in their leaf nodes, which allocates more space for insertions. This significantly reduces the likelihood of node splits.
Notably, \lisa\ exhibits an extremely low split ratio. It is visually close to zero in Figure~\ref{fig:node_splits_ratio}, though a small number of splits still occur.

\begin{figure}[h]
  \centering
  \includegraphics[width=0.45\textwidth]{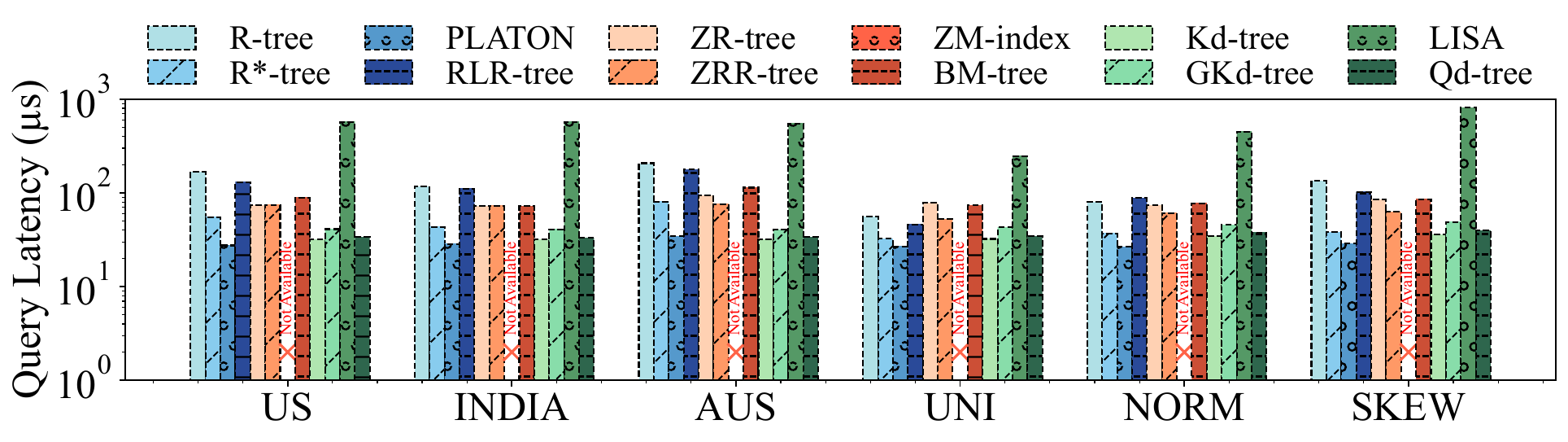}
  \caption{Query latency during insertion (write-heavy)}
  \label{fig:write_heavy_query_time}
\end{figure}

\subsubsection{Query Latency}
Figure~\ref{fig:write_heavy_query_time} shows query latency for write-heavy workload. Due to space constraints, read-heavy results
are provided in the supplementary materials.
We observe that:

\stepcounter{observation}
\noindent\textbf{O\theobservation: 
RLESIs exhibit comparable query latency under heavy insertions.
}
\rlrtree\ and \qdtree\ slightly outperform \rtree\ and \kdtree, respectively, while \bmtree\ performs similarly to \zrtree. This contrasts with the results of point queries alone (cf. Figure~\ref{fig:point_query}).
\lisa\ inserts more points into each leaf node, which increases the search cost at the leaf level, even though its index traversal remains efficient. 
Other SP-based indices show low query latency as insertion-related workloads use only one-tenth of the dataset, which reduces the index height for SP-based indices, thereby lowering query latency. 
In contrast, the height of DP- and MP-based indices remains largely unchanged, resulting in minimal latency variation.

\begin{figure}[h]
  \centering
\includegraphics[width=0.45\textwidth]{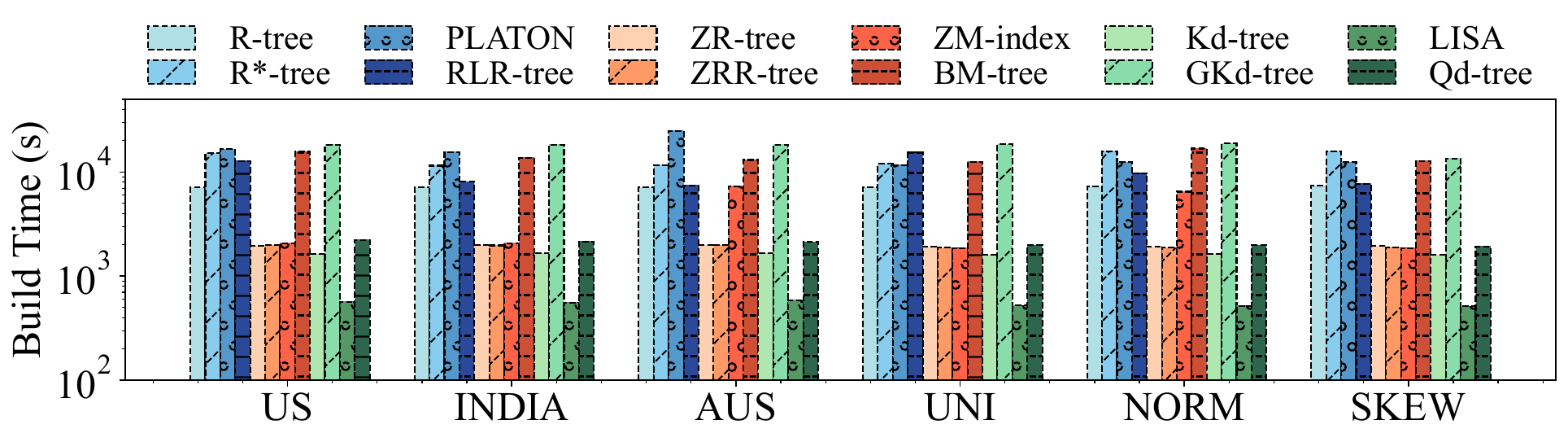}
  \caption{Index build time (training time included)} 
  \label{fig:build}
\end{figure}

\vspace{0.5em}
\begin{table}[h!]
\small
\centering
\caption{Training ``+'' building time of RLESIs (minutes).}
\resizebox{\linewidth}{!}{%
\begin{tabular}{c|c|c|c|c|c|c}
\toprule
Index & US & INDIA & AUS & UNI & NORM & SKEW \\ 
\midrule
\midrule
\rlrtree\ & 94 + 117 & 18 + 116 & 4 + 119 & 140 + 118 & 45 + 117 & 9 + 119 \\ \hline 
\qdtree\ & 6 + 30 & 5 + 29 & 4 + 29 & 5 + 28 & 4 + 27 & 4 + 27 \\ \hline
\bmtree\ & 257 + 3 & 226 + 2 & 214 + 2 & 206 + 2 & 277 + 2 & 209 + 2 \\ 
\bottomrule
\end{tabular}
}
\label{tab:train_build_time}
\end{table}

\subsection{Index Building Study}\label{subsec:index_build}
Figure~\ref{fig:build} shows total build time, and Table~\ref{tab:train_build_time} breaks down the training and building costs for RLESIs. We observe that:
\stepcounter{observation}
\noindent\textbf{O\theobservation: 
RLESIs, \rstar, \platon, and \rstar\ incur high index building costs.}
RLESIs are time-consuming to build, due to the expensive training process. 
For example,
\rlrtree\ incurs additional overhead from its training process, while both \platon\ and \bmtree\ utilize Monte Carlo tree search (MCTS) for index training, which has high computational complexity. Notably, the time cost for parameter tuning in RLESIs, exceeding 30 hours per dataset for each index, is omitted here.
For \rstar, the reinsertion process
is particularly time-intensive during index building. Similarly, \kdgreedy\ involves computationally expensive operations to determine the optimal partition locations~\cite{qdtree}.

\subsubsection{Index Size and Height}
Figure~\ref{fig:index_size} shows the index sizes, and Figure~\ref{fig:tree_height} shows the index heights. We observe that:

\begin{figure}[h]
  \centering
\includegraphics[width=0.45\textwidth]{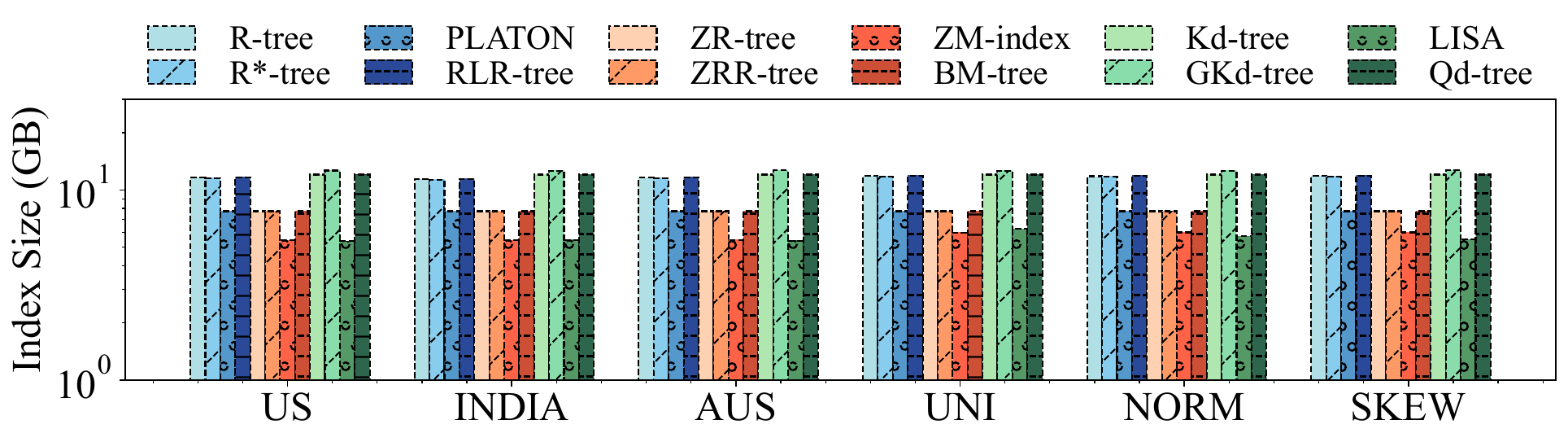}
  \caption{Index size} 
  \label{fig:index_size}
\end{figure}

\begin{figure}[h]
  \centering
\includegraphics[width=0.45\textwidth]{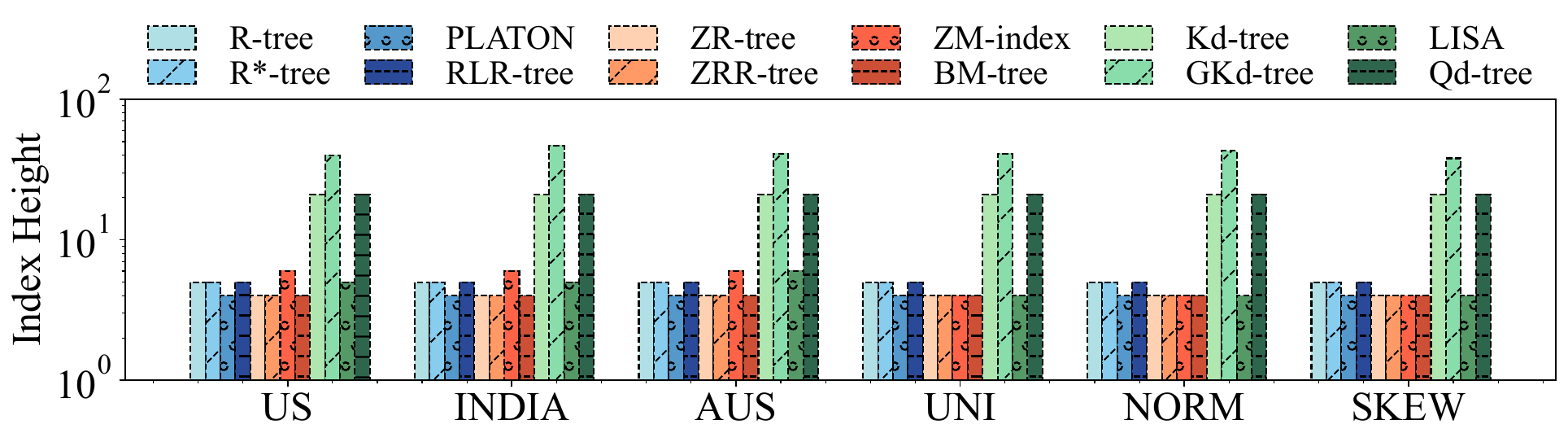}
  \caption{Index height} 
  \label{fig:tree_height}
\end{figure}
\stepcounter{observation}
\noindent\textbf{O\theobservation: The MP-based RLESI, \bmtree, is the most space-efficient among all indices, except for \lisa\ and \zmindex.}
This is because MP-based indices, including \bmtree, tightly pack data in leaf nodes, with only the last node potentially partially filled. This leads to lower tree height and smaller index size, as shown in Figure~\ref{fig:tree_height}. 
In contrast, DP-based RLESIs like \rlrtree\ use a lower fill ratio (0.4 in this study), leading to only 67\%–73\% node utilization. SP-based RLESIs, such as \qdtree, also incur higher space costs, as internal nodes have fixed fanout and leaf nodes are not always fully utilized (around 95\%). 
Notably, LSIs like \lisa\ and \zmindex\ save space by storing more data in each leaf and using linear models instead of tree-based inner structures, thus reducing the number of intermediate nodes.

 \vspace{0.5em}
\begin{table}[h!]
\small
\centering
\caption{The number of node adjustments (million).}
\resizebox{\linewidth}{!}{%
\begin{tabular}{c|c|c|c|c|c|c}
\toprule
Index & US & INDIA & AUS & UNI & NORM & SKEW \\ 
\midrule
\midrule
\rtree\ & 6.5  & 6.4 & 6.2 & 7.8 & 2.6 & 7.4 \\ \hline 
\rstar\ & 55.3 & 55.8 & 55.3 & 56.8 & 19.3 & 59.0 \\ \hline
\rlrtree\ & 7.4 & 7.1 & 7.1 & 8.5 & 8.4 & 8.9 \\ 
\bottomrule
\end{tabular}
}
\label{tab:node_adjustments}
\end{table}

\begin{figure}[h]
  \centering
\subfloat[Range query~\label{fig:range_query_time_varying_cardinality}]   {
\includegraphics[width=0.45\textwidth]{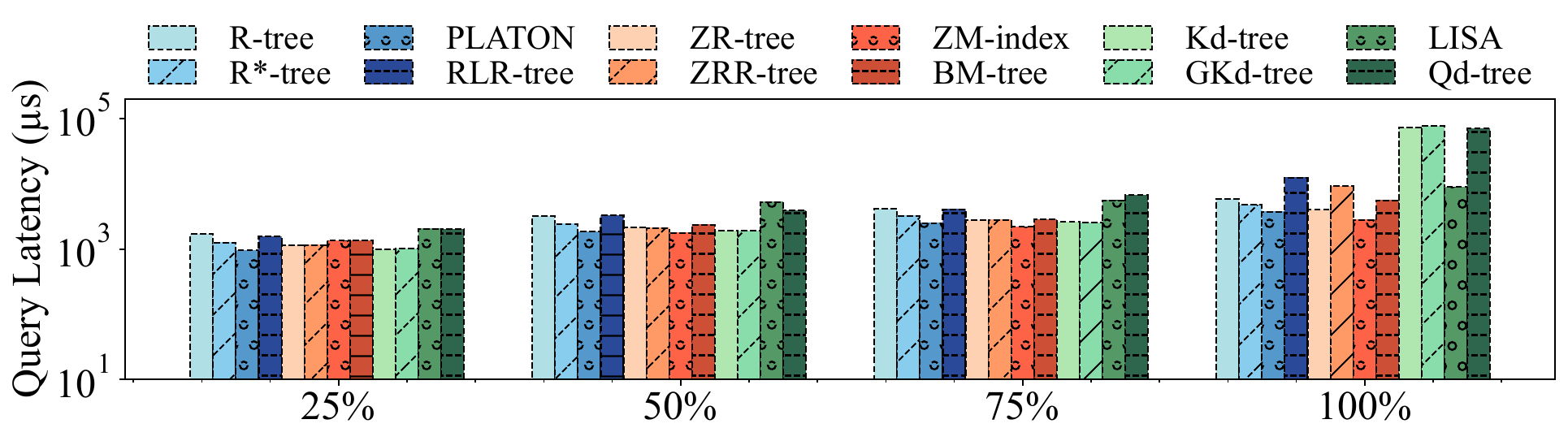}
}\\
\vspace{-1em}
\subfloat[Point query~\label{fig:point_query_time_varying_cardinality}]   {
\includegraphics[width=0.45\textwidth]{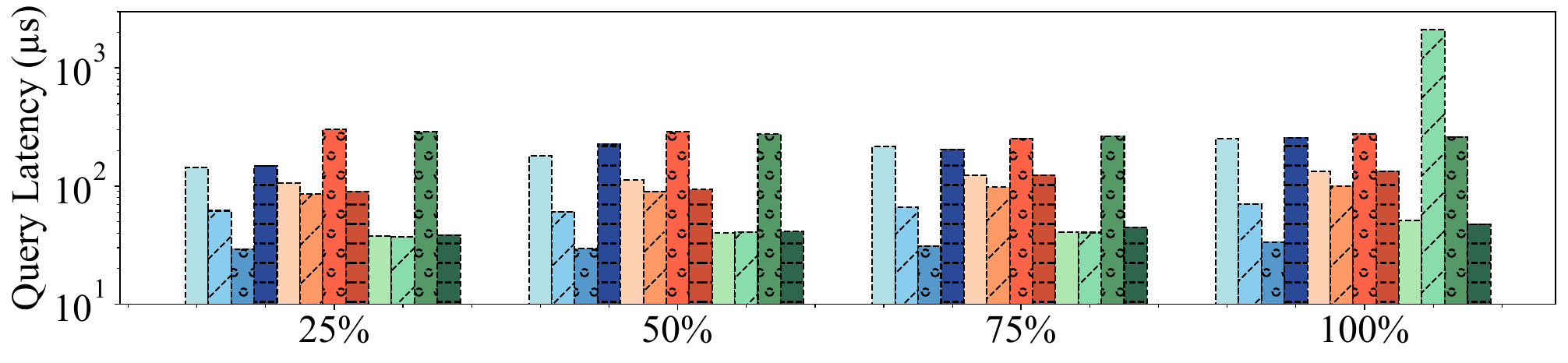}
}\\
\vspace{-1em}
\subfloat[$k$NN query~\label{fig:knn_query_time_varying_cardinality}]   {
\includegraphics[width=0.45\textwidth]{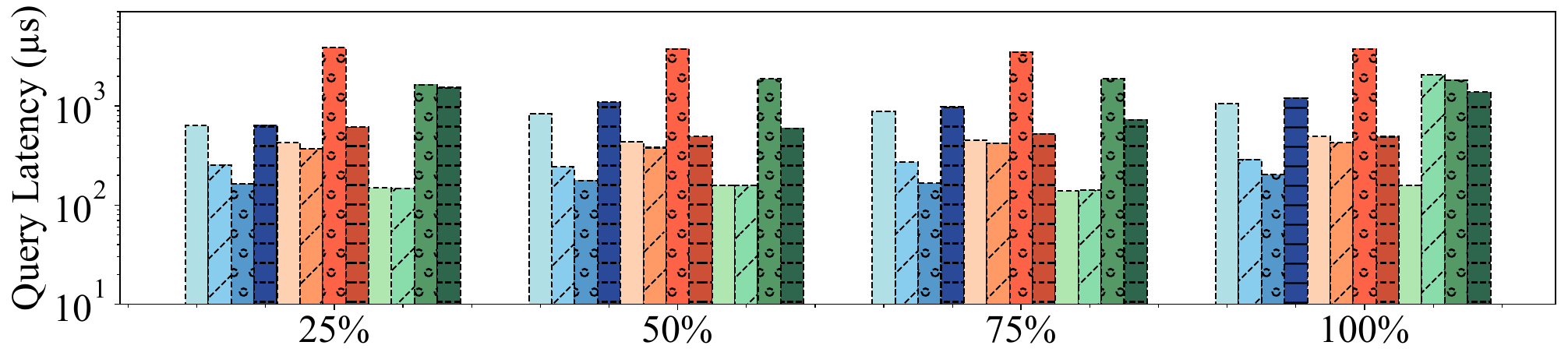}
}
  \caption{Varying dataset cardinality} 
  \label{fig:varying_cardinality}
\end{figure}

\begin{figure*}[h]
  \centering
  \includegraphics[width=0.89\textwidth]{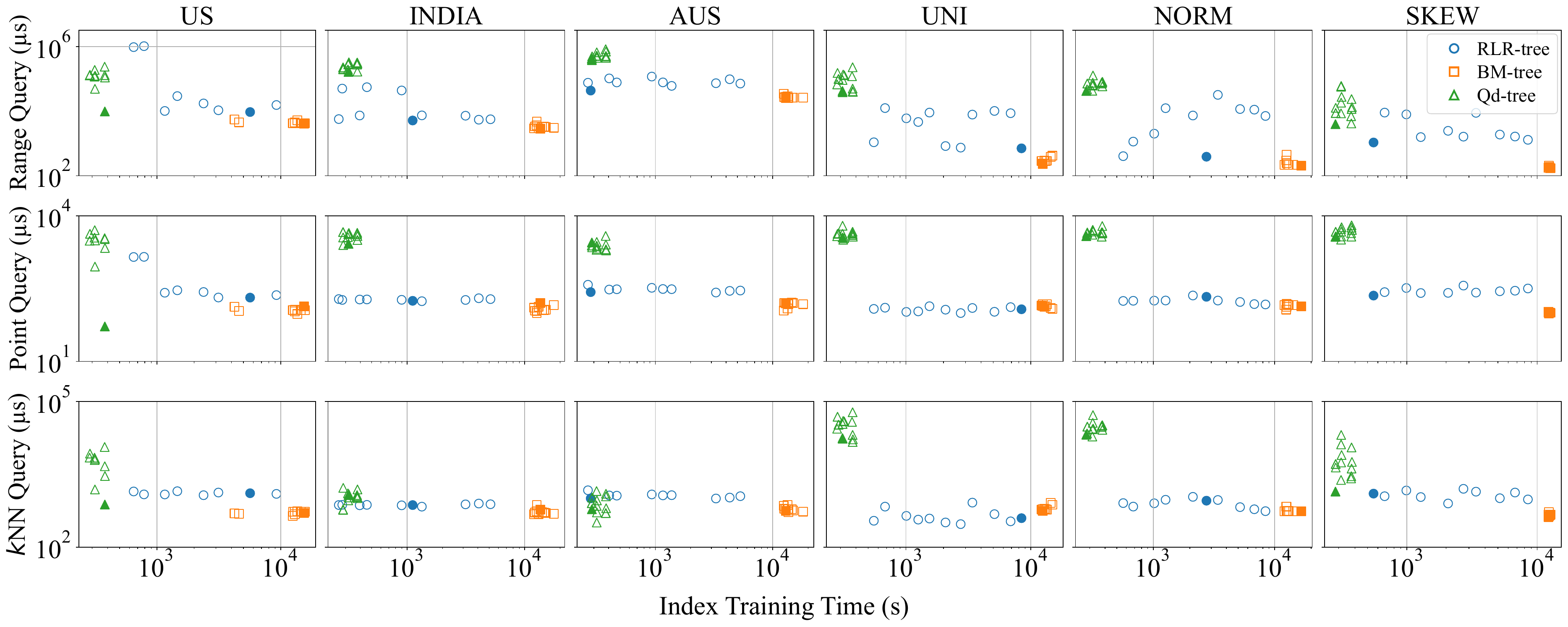}
  \caption{Query latency for range, point, and $k$NN queries vs. index training time across varying training configurations.}
  \label{fig:all_query_time_build_time}
\end{figure*}

\stepcounter{observation}
\noindent\textbf{O\theobservation: SP-based indices exhibit the highest tree height, except for \lisa.}
This is because binary partitioning limits each internal node to two children, naturally leading to deeper trees. In contrast, MP-based indices such as \bmtree\ achieve the lowest tree height by fully packing nodes.
However, tree height is not a direct indicator of performance, as it reflects a trade-off between depth and fanout. Although \lisa\ appears shallower, it does not follow binary partitioning like typical SP-based indices and is used here as a representative LSI in the absence of a binary-partitioned learned index.

\subsubsection{Node Adjustment in Building}
Table~\ref{tab:node_adjustments} presents node adjustments for three DP-based indices.
\platon, SP-based, and MP-based indices are excluded as they write sequentially on storage without adjustments.
We observe that:

\stepcounter{observation}
\noindent\textbf{O\theobservation: \rstar\ requires significantly more splits compared to \rlrtree, while \rlrtree\ requires only slightly more splits than \rtree.}
As previously mentioned, \rstar\ employs a reinsertion strategy, reinserting a portion of the entries when a node overflows. In contrast, \rtree\ and \rlrtree\ only adjust nodes during node splits.
Additionally, \rstar\ is more sensitive to skewed data, as evidenced in Table~\ref{tab:node_adjustments}, where it exhibits the highest number of node adjustments on the SKEW dataset.
\rlrtree\ shows more node adjustments than \rtree\ because it optimizes node splits to minimize node size.

\subsection{Cardinality Variation Study}\label{subsec:cardinality_variation} 
Figure~\ref{fig:varying_cardinality} shows query latency and index building cost as the cardinality of the US dataset varies.
We observe that:

\stepcounter{observation}
\noindent\textbf{O\theobservation: RLESIs exhibit stable query latency with increasing dataset size, consistent with trends across most indices, highlighting strong scalability for all query types.}
This stability reflects the scalability of the indices, which preserve performance trends as data volume grows. These results highlight the robustness of RLESIs and LSIs in large-scale settings, consistently delivering reliable query performance across different query types.

\subsection{Parameter Tuning Study}\label{subsec:rlesi_tuning} 
We tune all RLESIs using the configurations listed in 
Table~\ref{tab:configs}. While the search space remains limited, the chosen parameters have a substantial influence on training overhead and are sufficient to balance tuning efficiency against query performance.
We use a grid-based search with query and build cost limits as it is simple to implement and easy to reuse, given the small number of parameters.

\vspace{0.5em}
\begin{table}[h!]
\footnotesize
\centering
\caption{Configurations for RLESIs.}
\resizebox{\linewidth}{!}{%
\begin{tabular}{c|l}
\toprule
Index & Configurations \\ 
\midrule
\midrule
\rlrtree\ & \{epoch:[6,8,10,12,14], sample:[10k, 20k, 40k, 80k]\} \\ \hline 
\qdtree\ & \{query:[200, 400, 800], sample:[10k, 20k, 40k, 80k, 160k]\} \\ \hline
\bmtree\ & \{height:[8, 10, 12], sample:[10k, 20k, 40k, 80k]\} \\ 
\bottomrule
\end{tabular}
}
\label{tab:configs}
\end{table}

We measure parameter effectiveness using latency on the default range query workload, since all RLESIs define their training objectives on range access. 
Once chosen, the parameters are reused across all workloads on the same dataset, rather than re-tuned for each workload.
Figure~\ref{fig:all_query_time_build_time} illustrates the relationship between training time and query latency for range, point, and $k$NN queries across different datasets. 
Configurations achieving the lowest range query latency are highlighted with filled markers.
Overall, the tuning process requires a total of approximately $30 \times 3 \times 6$ hours (30 hours per baseline across 3 RLESIs and 6 datasets).
We observe that:

\stepcounter{observation}
\noindent\textbf{O\theobservation: Parameter tuning improves range query performance by up to two orders of magnitude.}
On the US dataset, the query latency for \rlrtree\ shows a 120$\times$ improvement between the best and worst-performing configurations. Significant enhancements are also observed across other datasets for \rlrtree. Similarly, for \qdtree, tuning parameters on the US dataset achieves up to a 28$\times$ reduction in range query latency, with approximately a 10$\times$ improvement on the AUS and SKEW datasets. In contrast, \bmtree\ shows limited improvement, likely due to restricting the sample size to minimize training time. Larger sample sizes are avoided to prevent excessive training costs.

\stepcounter{observation}
\noindent\textbf{O\theobservation: \qdtree\ exhibits consistent training costs, \rlrtree\ shows variability, while \bmtree\ is consistently slow but stable.}
\qdtree\ uses a mirrored \kdtree\ structure for learning, where the fixed two-partition splitting minimizes the impact of parameter changes. Conversely, \rlrtree\ explores multiple partitioning strategies during R-tree building, leading to higher and more variable training costs. This sensitivity to parameter tuning results in scattered performance. In contrast, \bmtree\ incurs consistently high training costs as it builds intermediate indices and uses real queries to compute rewards, with parameter changes having minimal effect.

\stepcounter{observation}
\noindent\textbf{O\theobservation: Different query types require tailored configurations, and no single RLESI excels across all query types, even with tuning.}
\qdtree\ benefits from efficient training and delivers strong performance on the US dataset. \rlrtree\ demonstrates significant potential through tuning in expanded parameter space, achieving competitive performance for point and $k$NN queries across most datasets. \bmtree, despite its longer training time, offers consistent efficiency and stability across all query types.

 \begin{figure}[h]
  \centering
  \includegraphics[width=0.47\textwidth]{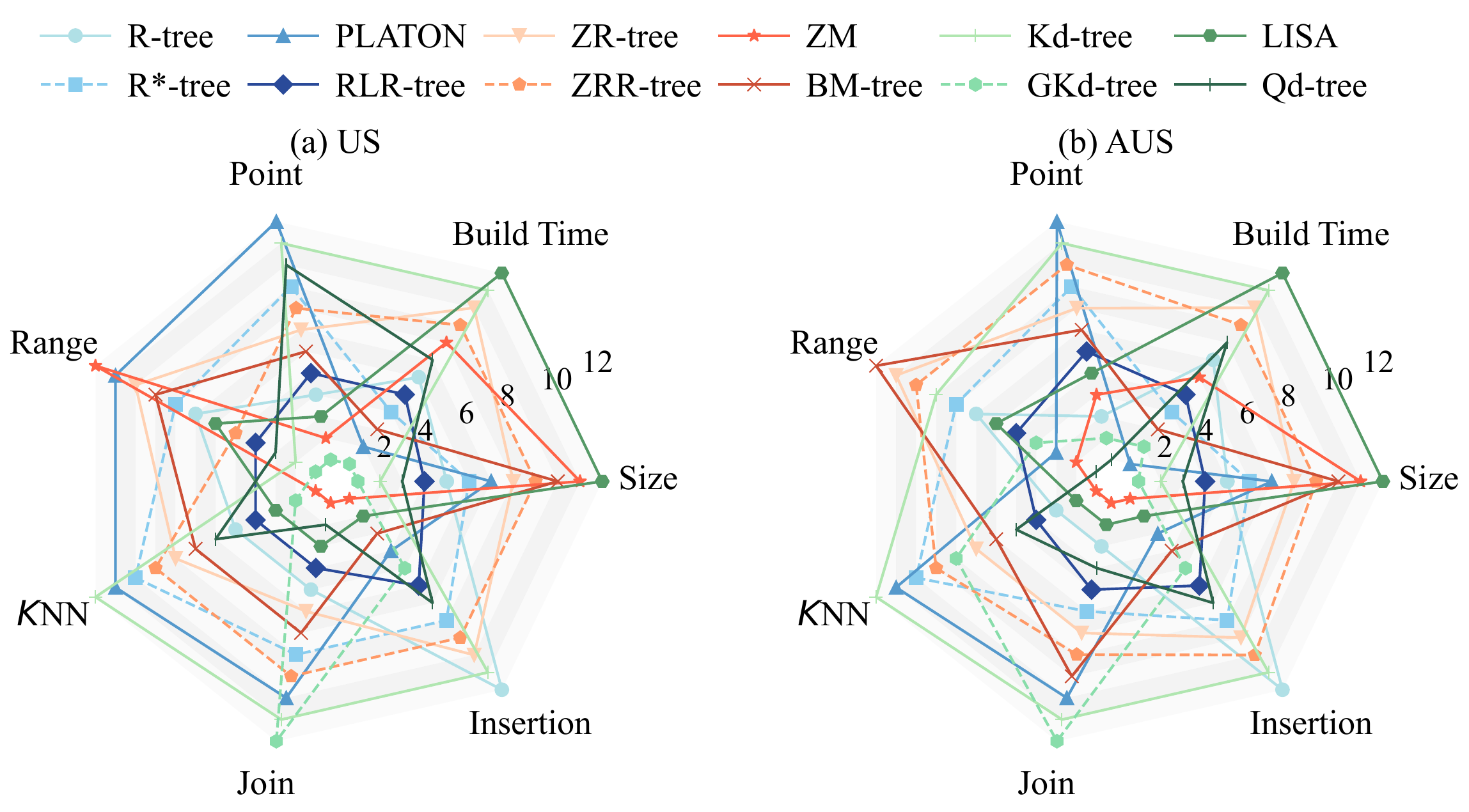}
  \caption{Comparison across five workloads, build time, and index size on US and AUS datasets. Each axis shows a rank-based score (higher is better).
  } 
  \label{fig:spider}
\end{figure}

\section{Insights and Limitations}\label{sec:insights}
\subsection{
Summary of Overall Evaluation}
Figure~\ref{fig:spider} summarizes the overall performance of all baselines on two 
representative datasets, US and AUS, evaluated across seven metrics: index size, build cost, query latency for four query types, and insertion performance.  Among all indices, \lisa\ is the most efficient in both build time and index size. In contrast, RLESIs generally incur higher build costs due to training overhead and the need for hyperparameter tuning to avoid suboptimal configurations. In terms of query performance, \platon, \zmindex, and \kdtree\ demonstrate strong performance for point, range, $k$NN, and join queries, respectively. However, RLESIs show unstable performance for point and $k$NN queries, as they are trained only on range queries and lack generalization. For insertion latency, \rtree\ is the most efficient. Its leaf nodes are underutilized compared to MP-based indices, and it avoids the reinsertion overhead seen in \rstar. In contrast, RLESIs suffer from additional overhead during insertion due to model inference.

\subsection{Challenges and Limitations of RLESIs}
\textbf{Challenges.}
As discussed in Section~\ref{subsec:rlesi_tuning},
training RLESIs is costly, as each parameter combination requires a full 
\emph{train $\rightarrow$ build $\rightarrow$ evaluate} cycle. 
Here, we train on sampled data, but build and evaluate the index on the full dataset to ensure accuracy.
This mismatch means models often optimize policies on artificial geometries that fail to capture true data locality. 
For example, in the full dataset, 100 spatially close points may be spread across multiple pages, whereas in a sparse sample the same 100 points can fit on a single page.
Therefore, the trained models do not capture data locality well.

\textbf{Limitations.}
Within the cycle, the costs of index build and evaluating are fixed, while training varies. From our tuning study in Section~\ref{subsec:tuning}, 
we know that parameter selection leads to cost differences. Beyond that, reward calculation also affects training overhead. For instance, \bmtree’s reward mechanism repeatedly evaluates intermediate partitions, resulting in high training cost (Figure~\ref{fig:all_query_time_build_time}).

\subsection{Opportunities for Improvement}
\begin{figure}[h]
  \centering
  \includegraphics[width=0.47\textwidth]{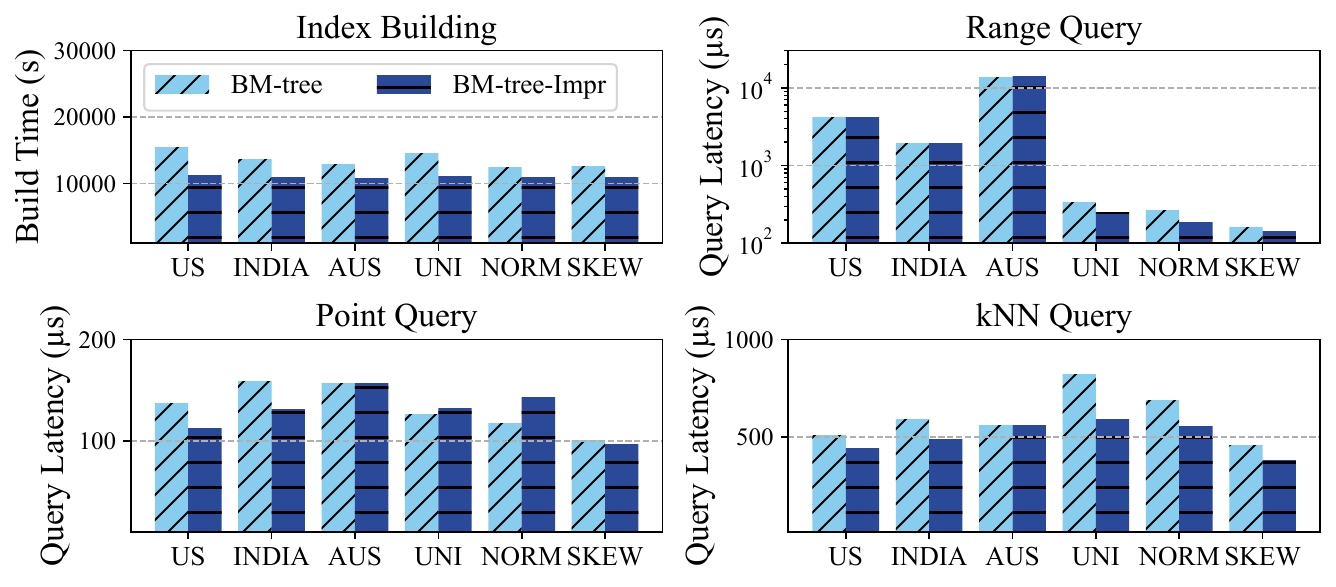}
  \caption{Comparison between \bmtree\ and BM-tree-Impr.} 
  \label{fig:bmtree_improved_time}
\end{figure}

To address the limitation, we enhance the training process using advanced reward calculations. Building on the approach in~\cite{LBMC}, we utilize cost estimations to optimize \bmtree's performance by refining its original cost function, which we refer to as BM-tree-Impr. As shown in Figure~\ref{fig:bmtree_improved_time}, build time improves by 12\% on the NORM and SKEW datasets, and up to 27\% on the US dataset. 

We also find two promising directions to reduce training cost: \textit{incremental learning} and the use of \textit{pre-trained models}. 
Incremental learning allows a model to update continuously as new data arrives, without retraining from scratch, while pre-trained models provide efficient initialization that reduces the need for long training. 
In the context of RLESIs, combining these approaches would allow the RL policy to adapt to new insertions and workload shifts by fine-tuning only the affected parts, making them more practical in dynamic settings.

\renewcommand{\labelenumi}{\arabic{enumi}.}

\section{Conclusions}\label{sec:conclusion}
This paper presents the first unified benchmark for evaluating RL-enhanced spatial indices (RLESIs) alongside traditional, advanced, and learned spatial indices. Our framework decouples index training from index building via two modular components, enabling extensible and reproducible evaluation.

To ensure fair comparison, we implement parameter tuning for RLESIs and evaluate 12 representative indices across six datasets and seven workloads using a comprehensive set of metrics. We find that while RLESIs can reduce query latency with proper tuning, they do not consistently outperform LSIs or advanced variants in query efficiency or index building time.

We identify key challenges and limitations of RLESIs, including high training overhead and sensitivity to parameter choices. We also address the high training cost for \bmtree\ by exploring cost-based reward functions and list potential opportunities to accelerate the RLESI tuning efficiency.

Overall, our study suggests that RLESIs are viable for systems requiring backward compatibility, but their practical adoption remains limited by high tuning costs and inconsistent performance across workloads.

\section*{Acknowledgments}


Zhifeng Bao is supported in part by the Australian Research Council (ARC) FT240100832, DP240101211.
Renata Borovica-Gajic is supported in part by the ARC Discovery Early Career Researcher Award DE230100366 and the Google Foundational Science 2025 fund.

\section{AI-Generated Content Acknowledgement}
The authors used ChatGPT in English language editing to improve the fluency, grammar, and clarity of text originally drafted by the authors. In addition, ChatGPT was also employed during code development to support basic syntax and implementation-related tasks. No experimental results or original research contributions were generated solely by AI tools. The authors take full responsibility for the accuracy, originality, and integrity of all content presented in this paper.

\bibliographystyle{IEEEtran}
\balance
\bibliography{references}

@inproceedings{R_tree_join,
author = {Brinkhoff, Thomas and Kriegel, Hans-Peter and Seeger, Bernhard},
title = {Efficient processing of spatial joins using R-trees},
year = {1993},
booktitle = {SIGMOD},
pages = {237–246},
numpages = {10},
}

@inproceedings{qdtree,
  author    = {Yang, Zongheng and Chandramouli, Badrish and Wang, Chi and Gehrke, Johannes and Li, Yinan and Minhas, Umar Farooq and Larson, Per-{\r A}ke and Kossmann, Donald and Acharya, Rajeev},
  title     = {Qd-Tree: Learning Data Layouts for Big Data Analytics},
  year      = {2020},
  booktitle = {SIGMOD},
  pages     = {193--208}
}

@inproceedings{ALEX,
  author    = {Ding, Jialin and Minhas, Umar Farooq and Yu, Jia and Wang, Chi and Do, Jaeyoung and Li, Yinan and Zhang, Hantian and Chandramouli, Badrish and Gehrke, Johannes and Kossmann, Donald and Lomet, David and Kraska, Tim},
  title     = {{ALEX}: An Updatable Adaptive Learned Index},
  year      = {2020},
  booktitle = {SIGMOD},
  pages     = {969--984}
}

@inproceedings{LISA,
  author    = {Pengfei Li and
               Hua Lu and
               Qian Zheng and
               Long Yang and
               Gang Pan},
  title     = {{LISA:} {A} Learned Index Structure for Spatial Data},
  booktitle = {SIGMOD},
  pages     = {2119-2133},
  year      = {2020}
}

@inproceedings{ZM,
  author    = {Haixin Wang and Xiaoyi Fu and Jianliang Xu and Hua  Lu},
  booktitle = {MDM},
  title     = {Learned Index for Spatial Queries},
  year      = {2019},
  pages     = {569-574}
}

@inproceedings{Flood,
  author    = {Vikram Nathan and
               Jialin Ding and
               Mohammad Alizadeh and
               Tim Kraska},
  title     = {Learning Multi-Dimensional Indexes},
  booktitle = {SIGMOD},
  pages     = {985-1000},
  year      = {2020}
}

@article{learned_spatial_indexes,
  author  = {Varun Pandey and
             Alexander van Renen and
             Andreas Kipf and
             Ibrahim Sabek and
             Jialin Ding and
             Alfons Kemper},
  title   = {The Case for Learned Spatial Indexes},
  journal = {CoRR},
  volume  = {abs/2008.10349},
  year    = {2020}
}

@inproceedings{RMI,
  author    = {Kraska, Tim and Beutel, Alex and Chi, Ed H. and Dean, Jeffrey and Polyzotis, Neoklis},
  title     = {The Case for Learned Index Structures},
  booktitle = {SIGMOD},
  year      = {2018},
  pages     = {489--504}
}

@inproceedings{learnedmultiindexNeurIPS,
  year      = { 2019 },
  booktitle = { NeurIPS 2019 Workshop on Machine Learning for Systems },
  title     = { Learning Multi-dimensional Indexes },
  author    = { Vikram Nathan and
               Jialin Ding and
               Mohammad Alizadeh and
               Tim Kraska }
}

@article{sosd,
  author       = {Ryan Marcus and
                  Andreas Kipf and
                  Alexander van Renen and
                  Mihail Stoian and
                  Sanchit Misra and
                  Alfons Kemper and
                  Thomas Neumann and
                  Tim Kraska},
  title        = {Benchmarking Learned Indexes},
  journal      = {PVLDB},
  volume       = {14},
  number       = {1},
  pages        = {1--13},
  year         = {2020},
}

@article{learnedbench,
      title={How Good Are Multi-dimensional Learned Indices? An Experimental Survey}, 
      author={Qiyu Liu and Maocheng Li and Yuxiang Zeng and Yanyan Shen and Lei Chen},
      year={2024},
      journal = {CoRR},
      volume  = {abs/2405.05536},
}

@article{are_learned_ready,
author = {Wongkham, Chaichon and Lu, Baotong and Liu, Chris and Zhong, Zhicong and Lo, Eric and Wang, Tianzheng},
title = {Are updatable learned indexes ready?},
year = {2022},
volume = {15},
number = {11},
journal = {PVLDB},
pages = {3004–3017},
}

@article{tsunami,
  author  = {Jialin Ding and
             Vikram Nathan and
             Mohammad Alizadeh and
             Tim Kraska},
  title   = {Tsunami: {A} Learned Multi-dimensional Index for Correlated Data and
             Skewed Workloads},
  year    = {2020},
  volume  = {14},
  number  = {2},
  journal = {PVLDB},
  pages   = {74–86}
}

@article{RSMI,
  author  = {Jianzhong Qi and
             Guanli Liu and
             Christian S. Jensen and
             Lars Kulik},
  title   = {Effectively Learning Spatial Indices},
  journal = {PVLDB},
  volume  = {13},
  number  = {11},
  pages   = {2341-2354},
  year    = {2020}
}

@article{RLR-tree,
  author       = {Tu Gu and
                  Kaiyu Feng and
                  Gao Cong and
                  Cheng Long and
                  Zheng Wang and
                  Sheng Wang},
  title        = {{The RLR-Tree: {A} Reinforcement Learning Based R-Tree for Spatial
                  Data}},
  journal      = {{PACMMOD}},
  volume       = {1},
  number       = {1},
  pages        = {63:1--63:26},
  year         = {2023},
}

@article{hilbert,
  author  = {Jonathan K Lawder and
             Peter J H King},
  title   = {Querying multi-dimensional data index using the Hilbert 
             space-filling curve},
  journal = {SIGMOD Rec.},
  volume  = {30},
  pages   = {19-24},
  number  = {1},
  year    = {2001}
}

@inproceedings{Zcurve,
  author    = {Orenstein, Jack A. and Merrett, T. H.},
  title     = {A Class of Data Structures for Associative Searching},
  booktitle = {PODS},
  year      = {1984},
  pages     = {181--190}
}

@inproceedings{ZRtree,
author = {Orenstein, Jack A.},
title = {Spatial Query Processing in an Object-Oriented Database System},
year = {1986},
booktitle = {SIGMOD},
pages = {326--336},
numpages = {11},
}

@article{quadtree,
  author  = {Raphael FinkelJon and Louis Bentley},
  title   = {Quad Trees: {A} Data Structure for Retrieval on Composite Keys},
  journal = {{Acta Informatica}},
year = {1974},
  volume  = {4},
  number  = {1},
  pages   = {1-9}
}

@inproceedings{HRtree,
  author    = {Kamel, Ibrahim and Faloutsos, Christos},
  title     = {Hilbert {R}-tree: An Improved {R}-tree Using Fractals},
  booktitle = {PVLDB},
  year      = {1994},
  pages     = {500--509}
}

@inproceedings{STR,
  author    = {Scott T. Leutenegger and
               J. M. Edgington and
               Mario A. L{\'{o}}pez},
  title     = {{STR}: {A} Simple and Efficient Algorithm for {R-T}ree Packing},
  booktitle = {{Proceedings of the IEEE International Conference on Data Engineering (ICDE)}},
  pages     = {497--506},
  year      = {1997}
}

@article{kdtree,
  author  = {Jon Louis Bentley},
  title   = {Multidimensional Binary Search Trees Used for Associative Searching},
  journal = {Communications of the ACM},
  month   = {},
  number  = {9},
  year    = {1975},
  volume  = {18},
  pages   = {509-517}
}

@article{PR_tree,
  author  = {Arge, Lars and Berg, Mark De and Haverkort, Herman and Yi, Ke},
  title   = {The {Priority R-tree}: A Practically Efficient and Worst-case Optimal {R-tree}},
  journal = {{ACM Transactions on Algorithms}},
  volume  = {4},
  number  = {1},
  year    = {2008},
  pages   = {9:1--9:30}
}

@inproceedings{rtree,
  author    = {Antonin Guttman},
  title     = {{R}-trees: A Dynamic Index Structure for Spatial Searching},
  booktitle = {SIGMOD},
  pages     = {47-57},
  year      = {1984}
}

@inproceedings{revised_r_star_tree,
  author    = {Beckmann, Norbert and Seeger, Bernhard},
  title     = {A Revised {R$^*$}-tree in Comparison with Related Index Structures},
  booktitle = {SIGMOD},
  year      = {2009},
  pages     = {799--812}
}

@inproceedings{r_star_tree,
  author    = {Norbert Beckmann and
               Hans-Peter Kriegel and
               Ralf Schneider and
               Bernhard Seeger},
  title     = {The {R$^*$}-Tree: An Efficient and Robust Access Method for Points
               and Rectangles},
  booktitle = {SIGMOD},
  year      = {1990},
  pages     = {322-331}
}

@article{grid,
  author  = {Nievergelt, J. and Hinterberger, Hans and Sevcik, Kenneth C.},
  title   = {The {G}rid {F}ile: An Adaptable, Symmetric Multikey File Structure},
  journal = {TODS},
  volume  = {9},
  number  = {1},
  year    = {1984},
  pages   = {38--71}
}

@article{pgm,
  author  = {Paolo Ferragina and Giorgio Vinciguerra},
  title   = {The {PGM-index}: a fully-dynamic compressed learned index with provable worst-case bounds},
  year    = {2020},
  volume  = {13},
  number  = {8},
  pages   = {1162-1175},
  journal = {PVLDB}
}

@inproceedings{fiting_tree,
  author    = {Galakatos, Alex and Markovitch, Michael and Binnig, Carsten and Fonseca, Rodrigo and Kraska, Tim},
  title     = {FITing-Tree: A Data-Aware Index Structure},
  booktitle = {SIGMOD},
  pages     = {1189-1206},
  year      = {2019}
}

@article{BMTree,
  author    = {Jiangneng Li and Zheng Wang and Gao Cong and Cheng Long and Han Mao Kiah and Bin Cui},
  title     = {Towards Designing and Learning Piecewise Space-Filling Curves},
  journal   = {PVLDB},
  volume    = {16},
  number    = {9},
  pages     = {2158-2171},
  year      = {2023},
}

@article{LMSFC,
  author    = {Jian Gao and Xin Cao and Xin Yao and Gong Zhang and Wei Wang},
  title     = {LMSFC: A Novel Multidimensional Index based on Learned Monotonic Space Filling Curves},
  journal   = {PVLDB},
  volume    = {16},
  number    = {10},
  pages     = {2605-2617},
  year      = {2023},
}

@article{ULI_Evaluation,
author = {Lan, Hai and Bao, Zhifeng and Culpepper, J. Shane and Borovica-Gajic, Renata},
title = {Updatable Learned Indexes Meet Disk-Resident DBMS - From Evaluations to Design Choices},
year = {2023},
volume = {1},
number = {2},
journal = {Proceedings of the ACM on
Management of Data},
}

@inproceedings{LI_disk,
  author       = {Hai Lan and
                  Zhifeng Bao and
                  J. Shane Culpepper and
                  Renata Borovica{-}Gajic and
                  Yu Dong},
  title        = {A Fully On-Disk Updatable Learned Index},
  booktitle    = {40th {IEEE} International Conference on Data Engineering, {ICDE} 2024,
                  Utrecht, The Netherlands, May 13-16, 2024},
  pages        = {4856--4869},
  publisher    = {{IEEE}},
  year         = {2024},
  url          = {https://doi.org/10.1109/ICDE60146.2024.00369},
  doi          = {10.1109/ICDE60146.2024.00369},
  bibsource    = {dblp computer science bibliography, https://dblp.org}
}

@ARTICLE{Actor_critic,
  author={Grondman, Ivo and Busoniu, Lucian and Lopes, Gabriel A. D. and Babuska, Robert},
  journal={IEEE Transactions on Systems, Man, and Cybernetics, Part C (Applications and Reviews)}, 
  title={A Survey of Actor-Critic Reinforcement Learning: Standard and Natural Policy Gradients}, 
  year={2012},
  volume={42},
  number={6},
  pages={1291-1307},
}

@INPROCEEDINGS{AI_RTREE,
  author={Abdullah-Al-Mamun, Abdullah-Al- and Haider, Ch. Md. Rakin and Wang, Jianguo and Aref, Walid G.},
  booktitle={MDM}, 
  title={The “AI + R” - tree: An Instance-optimized R - tree}, 
  year={2022},
  pages={9-18}}

@InProceedings{SMBO,
author={Hutter, Frank
and Hoos, Holger H.
and Leyton-Brown, Kevin},
title={Sequential Model-Based Optimization for General Algorithm Configuration},
booktitle={International Conference on Learning and Intelligent Optimization},
year={2011},
pages={507--523},
}

@online{libspatialindex,
  author = {Marios Hadjieleftheriou},
  title = {libspatialindex},
  year = {2024},
  howpublished = {\url{https://github.com/libspatialindex/libspatialindex}},
  note = {Accessed: 2024-05-14}
}

@online{pytorch,
  author = {{PyTorch}},
  title  = {\url{https://pytorch.org}},
  year   = 2016,
  note   = {Accessed: 2023-05-31}
}

@online{btree,
  author = {{STX B+ Tree}},
  title  = {\url{https://panthema.net/2007/stx-btree}},
  year   = {2007},
  note   = {Accessed: 2023-05-31}
}

@online{Oracle,
  author = {{Oracle Manual}},
  title  = {\url{https://docs.oracle.com/en/database/oracle/oracle-database/23/spatl/creating-spatial-index.html}},
  year   = 2024,
  note   = {Accessed: 2024-12-27}
}

@online{MS_SQL,
  author = {{Microsoft SQL Document}},
  title  = {\url{https://learn.microsoft.com/en-us/sql/relational-databases/spatial/spatial-indexes-overview?view=sql-server-ver15}},
  year   = 2023,
  note   = {Accessed: 2024-12-27}
}

@online{PostGIS,
  author = {{PostGIS}},
  year   = {2024},
  title  = {\url{https://postgis.net/docs/manual-3.5/using_postgis_dbmanagement.html\#spgist_indexes}},
  note   = {Accessed: 2024-12-27}
}

@article{rank_space,
	author    = {Jianzhong Qi and
	Yufei Tao and
	Yanchuan Chang and
	Rui Zhang},
	title     = {Theoretically Optimal and Empirically Efficient {R}-trees with Strong
	Parallelizability},
	journal   = {{PVLDB}},
	volume    = {11},
	number    = {5},
	pages     = {621--634},
	year      = {2018}
}

@article{Packing_Rtrees_SFC,
author = {Qi, Jianzhong and Tao, Yufei and Chang, Yanchuan and Zhang, Rui},
title = {Packing R-trees with Space-filling Curves: Theoretical Optimality, Empirical Efficiency, and Bulk-loading Parallelizability},
year = {2020},
volume = {45},
number = {3},
journal = {TODS},
numpages = {47}
}

@inproceedings{ML-Index,
  author    = {Angjela Davitkova and
               Evica Milchevski and
               Sebastian Michel},
  title     = {The {ML-Index}: {A} Multidimensional, Learned Index for Point, Range,
               and Nearest-Neighbor Queries},
  booktitle = {EDBT},
  pages     = {407--410},
  year      = {2020}
}

@inproceedings{SPRIG,
author = {Zhang, Songnian and Ray, Suprio and Lu, Rongxing and Zheng, Yandong},
title = {{SPRIG}: A Learned Spatial Index for Range and {kNN} Queries},
booktitle = {SSTD},
pages = {96--105},
year = {2021},
}

@ARTICLE{idistancetods,
    author={H. V. Jagadish and Beng Chin Ooi and Kian-Lee Tan and Cui Yu and Rui Zhang},
    title={i{D}istance: An Adaptive {B$^+$}-tree Based Indexing Method for Nearest Neighbor Search},
    journal={TODS},
    volume={30},
    number={2},
    year={2005},
    pages={364-397}
}

@book{Rtree_book,
author = {Manolopoulos, Yannis and Nanopoulos, Alexandros and Papadopoulos, Apostolos N. and Theodoridis, Yannis},
title = {R-Trees: Theory and Applications},
year = {2005},
isbn = {1852339772},
publisher = {Springer Publishing Company, Incorporated},
}

@inproceedings{RadixSpline,
  author    = {Kipf, Andreas and Marcus, Ryan and an Renen, Alexander and Stoian, Mihail and Kemper, Alfons and Kraska, Tim and Neumann, Thomas},
  title     = {{RadixSpline}: A Single-Pass Learned Index},
  year      = {2020},
  booktitle = {{Proceedings of the International Workshop on Exploiting Artificial Intelligence Techniques for Data Management (aiDM)}},
  pages     = {5:1--5:5}
}

@article{LIPP,
author = {Wu, Jiacheng and Zhang, Yong and Chen, Shimin and Wang, Jin and Chen, Yu and Xing, Chunxiao},
title = {Updatable Learned Index with Precise Positions},
year = {2021},
volume = {14},
number = {8},
journal = {PVLDB},
pages = {1276–1288},
numpages = {13}
}

@online{osm_stats,
  author = {{OpenStreetMap Stats}},
  title = {\url{https://www.openstreetmap.org/stats/data_stats.html}},
  year = 2024,
  note = {Accessed: 2024-07-22}
}

@inproceedings{KDB,
 author = {Robinson, John T.},
 title = {The {K-D-B-tree}: A Search Structure for Large Multidimensional Dynamic Indexes},
 booktitle = {SIGMOD},
 year = {1981},
 pages = {10--18}
}

@article{LBMC,
    author = {Guanli Liu and Lars Kulik and Christian S. Jensen and Tianyi Li and Renata Borovica-Gajic and Jianzhong Qi},
    title = {Efficient Cost Modeling of Space-filling Curves},
      journal = {PVLDB},
      volume  = {17},
      number  = {13},
      pages   = {4773-4785},
      year    = {2024}
}

@article{Waffle,
author = {Moti, Moin Hussain and Simatis, Panagiotis and Papadias, Dimitris},
title = {Waffle: A Workload-Aware and Query-Sensitive Framework for Disk-Based Spatial Indexing},
year = {2022},
volume = {16},
journal = {PVLDB},
pages = {670–683}
}

@InProceedings{ACR-Tree,
author={Huang, Shuai
and Wang, Yong
and Li, Guoliang},
title={ACR-Tree: Constructing R-Trees Using Deep Reinforcement Learning},
booktitle={DASFAA},
year={2023},
publisher={Springer},
pages={80--96},
}

@article{PLATON,
author = {Yang, Jingyi and Cong, Gao},
title = {PLATON: Top-down R-tree Packing with Learned Partition Policy},
year = {2023},
issue_date = {December 2023},
publisher = {Association for Computing Machinery},
address = {New York, NY, USA},
volume = {1},
number = {4},
journal = {PACMMOD},
pages={1--26},
}

@article{Spatial-LS,
      title={Enhancing In-Memory Spatial Indexing with Learned Search}, 
      author={Varun Pandey and Alexander van Renen and Eleni Tzirita Zacharatou and Andreas Kipf and Ibrahim Sabek and Jialin Ding and Volker Markl and Alfons Kemper},
      journal = {CoRR},
      year={2023},
      volume={abs/2309.06354},
      archivePrefix={arXiv}
}

@article{CaseSpatial,
  author       = {Varun Pandey and
                  Alexander van Renen and
                  Andreas Kipf and
                  Ibrahim Sabek and
                  Jialin Ding and
                  Alfons Kemper},
  title        = {The Case for Learned Spatial Indexes},
journal = {CoRR},
  volume       = {abs/2008.10349},
  year         = {2020},
      archivePrefix={arXiv}
}

@INPROCEEDINGS{Benchmark_ls,
  author={Bindschaedler, Laurent and Kipf, Andreas and Kraska, Tim and Marcus, Ryan and Minhas, Umar Farooq},
  booktitle={ICDEW}, 
  title={Towards a Benchmark for Learned Systems}, 
  year={2021},
  volume={},
  number={},
  pages={127-133},
 }

\end{document}